\documentclass{desyproc}
\pdfoutput=1
\numberwithin{equation}{section}
\newcommand{\blue}[1]{\text{\textcolor{blue}{$\displaystyle #1$}}}
\newcommand{\green}[1]{\text{\textcolor[rgb]{0,0.5,0}{$\displaystyle #1$}}}
\newcommand{\magenta}[1]{\text{\textcolor{magenta}{$\displaystyle #1$}}}
\newcommand{\red}[1]{\text{\textcolor{red}{$\displaystyle #1$}}}
\newcommand{\brown}[1]{\text{\textcolor{brown}{$\displaystyle #1$}}}
\definecolor{brown}{rgb}{0.59,0.29,0.0}
\newcommand{\D}{\rlap{\hspace{1mm}/}D}
\newcommand{\DD}{\rlap{\hspace{1mm}/}\mathcal{D}}
\DeclareMathOperator{\Tr}{Tr}

\begin{document}
\title{Lectures on Soft-Collinear Effective Theory}

\author{{\slshape Andrey Grozin}\\[1ex]
Budker Institute of Nuclear Physics, Lavrentiev St.~11, Novosibirsk 630090, Russia}

\contribID{}

\confID{}  
\desyproc{}
\acronym{} 

\maketitle

\begin{abstract}
Introductory lectures on SCET
mainly following the first chapters of~\cite{BBF:15}.
\end{abstract}

\section{Introduction}
\label{S:Intro}

Soft-Collinear Effective Theory (SCET) has been first proposed as an effective field theory
describing inclusive $B$ meson decays with an energetic light quark in the final state
and exclusive $B$ decays with an energetic light hadron in the final state~\cite{B:00,B:01,B:02}.
Physics at short distances ($\sim 1/M_b$) is encapsulated in matching coefficients;
soft dynamics of the $b$ quark is described by Heavy Quark Effective Theory (HQET);
soft and collinear degrees of freedom of light quarks and gluons are described by SCET.
Speaking more precisely, there are two closely related effective theories, SCET-I and SCET-II.
Later SCET was successfully applied to numerous problems in hadron collider physics.
In this area there are at least two large nearly light-like momenta,
and hence several collinear regions (jets, initial hadrons).
Additional effective field theories were proposed for various classes of such problems.

These lectures are an elementary introduction to SCET.
A very nice textbook on SCET has recently appeared~\cite{BBF:15},
I follow the first chapters of this book.
There are also lecture notes~\cite{BS}.
I don't cite original papers here, citations can be found in~\cite{BBF:15,BS}.
SCET can be formulated in two different languages:
the coordinate space Lagrangian~\cite{B:02} or the label formalism~\cite{B:00,B:01}.
The book~\cite{BBF:15} mostly follows the first approach,
while the lectures~\cite{BS} -- the second one.
I shall not describe the label formalism here.

Let's first discuss what are effective field theories.
Many problems in quantum field theory contain several widely separated energy scales.
Suppose there is a high momentum scale $Q$
(and a short distance scale $1/Q$),
and we are interested in interactions of light particles ($m_i \ll Q$)
having small momenta ($p_i \ll Q$),
or in other words, physics at large distances $\gg 1/Q$.
The ratio of the characteristic soft scale to $Q$ is the small expansion parameter $\lambda$.
One of approaches to such problems is the method of regions for loop diagrams~\cite{BS:97}.
A brief introduction to this method is given in Sect.~\ref{S:Reg},
see~\cite{S:02} for more details.

Another approach is low energy (or large distance) effective field theories.
One can construct an effective Lagrangian containing only light fields.
Physics at small distances $\lesssim 1/Q$ produces
local interactions of these fields.
The Lagrangian contains all possible operators
(allowed by symmetries of the full theory).
In order to find coefficients of these operators matching is performed.
We calculate some scattering amplitudes in the full theory
and expands them in $\lambda$ up to some order;
then we calculate the same amplitudes in the effective theory,
equate the results and find the coefficients in the effective Lagrangian.
In other words, hard degrees of freedom are integrated out from the path integral
producing the soft effective action.

Similarly, operators of the full theory can be represented as $1/Q$ expansions
in terms of operators of the effective theory
with appropriate quantum numbers.
Coefficients in such expansions are also obtained by matching:
calculate some on-shell matrix elements in the full theory,
expand in $\lambda$, calculate them in the effective theory,
equate the results and find matching coefficients.

There is a close correspondence between effective theories and the method of regions.
Hard regions produce matching coefficients,
while soft ones are Feynman diagrams of the effective theory.
If one wants to consider just a single process described by a small number of diagrams,
it is rather easy to find many terms of its $\lambda$ expansion by the method of regions.
The effective field theory framework quickly becomes very complicated
with the growth of order of this expansion;
usually, only a few first terms can be considered.
On the other hand, an effective field theory is applicable to all processes at once.
It often allows one to derive various properties
which are valid to all orders of perturbation theory,
such as factorization theorems and renormalization group (RG) equations.
The leading terms of an effective Lagrangian can have some symmetry
which is not obvious in the full theory.

SCET is more complicated than simple effective theories because it is anisotropic%
\footnote{This is also true for NRQCD and pNRQCD.}.
Characteristic momentum scales (and distance scales) are different in different directions.
We discuss SCET for the scalar $\varphi^3$ field theory in Sect.~\ref{S:Scal}.
This simple example shows most typical features of SCET;
however, the coupling constant $g$ is dimensionful,
and the situation with resummation of all orders of perturbation theory
radically differs fro that in QCD
($\varphi^3$ in $d=6$ dimensional space--time has dimensionless $g$,
and is more like QCD).
QCD is a gauge theory,
and additional complications related to gauge invariance and Wilson lines appear;
these topics are discussed in Sect.~\ref{S:QCD}.

\section{Method of regions}
\label{S:Reg}

\begin{wrapfigure}{r}{0.3\textwidth}
\begin{center}
\begin{picture}(22,28)
\put(11,14){\makebox(0,0){\includegraphics{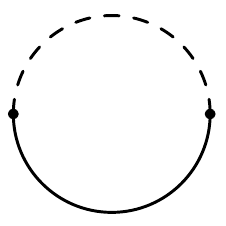}}}
\put(11,0){\makebox(0,0)[b]{$M$}}
\put(11,28){\makebox(0,0)[t]{$m$}}
\end{picture}
\end{center}
\caption{Vacuum diagram with two masses.}
\label{F:Reg}
\end{wrapfigure}
Let's consider the vacuum integral
\begin{equation*}
I = \int \frac{d^d k}{i\pi^{d/2}} \frac{1}{(M^2-k^2-i0) (m^2-k^2-i0)}
\end{equation*}
with two masses $M$ and $m$ (Fig.~\ref{F:Reg}) at $M\gg m$ and $d=2$.
It contains neither ultraviolet (UV) nor infrared (IR) divergences.
After Wick rotation to Euclidean momentum space it becomes
\begin{equation}
I = \int \frac{d^d k_E}{\pi^{d/2}} \frac{1}{(k_E^2+M^2) (k_E^2+m^2)}\,.
\label{Reg:Idef}
\end{equation}
Of course, in this simple example it is easy to obtain the exact solution.
We use partial fraction decomposition
\begin{equation*}
I = \frac{1}{M^2-m^2} \int \frac{d^dk_E}{\pi^{d/2}}
\left[ - \frac{1}{k_E^2+M^2} + \frac{1}{k_E^2+m^2} \right]\,,
\end{equation*}
\newpage
\begin{equation*}
\raisebox{-10.25mm}{\includegraphics{grozin_andrey_fig01.pdf}}
= \frac{1}{M^2-m^2} \Biggl[
- \raisebox{-10.25mm}{\includegraphics{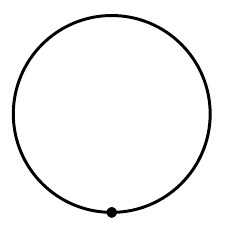}}
+ \raisebox{-10.25mm}{\includegraphics{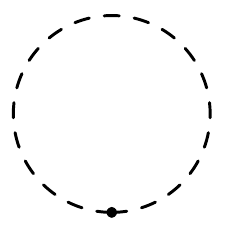}}
\Biggr]\,.
\end{equation*}
These two integrals, taken separately, diverge;
therefore, we use dimensional regularization ($d=2-2\varepsilon$) and obtain
\begin{equation*}
I = - \Gamma(\varepsilon) \frac{M^{-2\varepsilon} - m^{-2\varepsilon}}{M^2 - m^2}\,.
\end{equation*}
This result is finite at $\varepsilon\to0$:
\begin{equation}
I = \frac{\log(M^2/m^2)}{M^2 - m^2}
= \frac{1}{M^2} \log\frac{M^2}{m^2}
\left[ 1 + \frac{m^2}{M^2} + \frac{m^4}{M^4} + \cdots \right]\,.
\label{Reg:exact}
\end{equation}

It is easier to obtain this result using the prescription known as the method of regions~\cite{S:02}.
The integral~(\ref{Reg:Idef}) is written at the sum of contributions of two regions,
the hard one and the soft one:
\begin{equation*}
I = \red{I_h} + \magenta{I_s}\,.
\end{equation*}
In the hard region $\red{k_E\sim M}$;
in the soft one $\magenta{k_E\sim m}$.
The integrand is expanded to Taylor series
in accordance to these power counting rules in each region.
After that, the integral is taken over the full ($d$-dimensional) space.

In the hard region ($k_E\sim M$) we have
\begin{equation*}
\red{I_h} = \int \frac{d^dk_E}{\pi^{d/2}}
\red{T_h} \frac{1}{(k_E^2 + M^2) (k_E^2 + m^2)}\,,
\end{equation*}
where the operator $\red{T_h}$ expands the integrand in small parameter(s)
counting $k_E$ as a quantity of order $M$:
\begin{equation*}
\red{T_h} \frac{1}{(k_E^2 + M^2) (k_E^2 + m^2)}
= \frac{1}{k_E^2 + M^2} \frac{1}{k_E^2}
\left[ 1 - \frac{m^2}{k_E^2} + \frac{m^4}{k_E^4} - \cdots \right]\,.
\end{equation*}
Calculating the loop integrals, we arrive at
\begin{equation}
\red{I_h} = - \frac{M^{-2\varepsilon}}{M^2} \Gamma(\varepsilon)
\left[ 1 + \frac{m^2}{M^2} + \frac{m^4}{M^4} + \cdots \right]\,.
\label{Reg:hard}
\end{equation}
The result is a Taylor series in $m$.
Each loop integral is IR divergent;
it contains a single scale $M$,
and hence, by dimensions counting, is proportional to $M^{-2\varepsilon}$.

In the soft region ($k_E\sim m$) we have
\begin{equation*}
\magenta{I_s} = \int \frac{d^dk_E}{\pi^{d/2}}
\magenta{T_s} \frac{1}{(k_E^2 + M^2) (k_E^2 + m^2)}\,
\end{equation*}
where $\magenta{T_s}$ counts $k_E$ as a quantity of order $m$:
\begin{equation*}
\magenta{T_s} \frac{1}{(k_E^2 + M^2) (k_E^2 + m^2)}
= \frac{1}{M^2} \frac{1}{k_E^2 + m^2}
\left[ 1 - \frac{k_E^2}{M^2} + \frac{k_E^4}{M^4} - \cdots \right]\,,
\end{equation*}
and we obtain
\begin{equation}
\magenta{I_s} = \frac{m^{-2\varepsilon}}{M^2} \Gamma(\varepsilon)
\left[ 1 + \frac{m^2}{M^2} + \frac{m^4}{M^4} + \cdots \right]\,.
\label{Reg:soft}
\end{equation}
The result is a Taylor series in $1/M$.
Each loop integral is UV divergent;
it contains a single scale $m$,
and hence, by dimensions counting, is proportional to $m^{-2\varepsilon}$.

The sum of the hard contribution~(\ref{Reg:hard}) and the soft one~(\ref{Reg:soft})
produces the complete result~(\ref{Reg:exact}).
IR divergences in the hard region cancel UV divergences in the soft one.

In this simple case it is easy to prove this prescription~\cite{J:11}.
Let's introduce some boundary $\Lambda$ such that $m\ll\Lambda\ll M$,
and write
\begin{equation*}
I = \int_{k_E>\Lambda} \frac{d^dk_E}{\pi^{d/2}}
\frac{1}{(k_E^2 + M^2) (k_E^2 + m^2)}
+ \int_{k_E<\Lambda} \frac{d^dk_E}{\pi^{d/2}}
\frac{1}{(k_E^2 + M^2) (k_E^2 + m^2)}\,.
\end{equation*}
We may apply $T_h$ to the first integrand and $T_s$ to the second one:
\begin{equation*}
I = \int_{k_E>\Lambda} \frac{d^dk_E}{\pi^{d/2}}
T_h \frac{1}{(k_E^2 + M^2) (k_E^2 + m^2)}
+ \int_{k_E<\Lambda} \frac{d^dk_E}{\pi^{d/2}}
T_s \frac{1}{(k_E^2 + M^2) (k_E^2 + m^2)}\,.
\end{equation*}
For each of these two integrals, we add and subtract the integral
over the remaining part of the $k_E$ space:
\begin{equation*}
I = \red{I_h} + \magenta{I_s} - \Delta I\,,
\end{equation*}
where 
\begin{equation*}
\Delta I = \int_{k_E<\Lambda} \frac{d^dk_E}{\pi^{d/2}}
T_h \frac{1}{(k_E^2 + M^2) (k_E^2 + m^2)}
+ \int_{k_E>\Lambda} \frac{d^dk_E}{\pi^{d/2}}
T_s \frac{1}{(k_E^2 + M^2) (k_E^2 + m^2)}
\end{equation*}
are the integrals of the two expansions over the ``wrong'' parts of the $k_E$ space.
In these wrong regions, we may apply the other expansion in addition to the existing one:
\begin{equation*}
\Delta I = \int_{k_E<\Lambda} \frac{d^dk_E}{\pi^{d/2}}
T_s T_h \frac{1}{(k_E^2 + M^2) (k_E^2 + m^2)}
+ \int_{k_E>\Lambda} \frac{d^dk_E}{\pi^{d/2}}
T_h T_s \frac{1}{(k_E^2 + M^2) (k_E^2 + m^2)}\,.
\end{equation*}
But this is the integral over the full $k_E$ space:
\begin{equation*}
\Delta I = \int \frac{d^dk_E}{\pi^{d/2}}
T_s T_h \frac{1}{(k_E^2 + M^2) (k_E^2 + m^2)}\,,
\end{equation*}
because the Taylor expansion operators commute:
\begin{align*}
&T_s T_h \frac{1}{(k_E^2 + M^2) (k_E^2 + m^2)}
= T_h T_s \frac{1}{(k_E^2 + M^2) (k_E^2 + m^2)}\\
&{} = \frac{1}{M^2 k_E^2}
\left[ 1 - \frac{m^2}{k_E^2} + \frac{m^4}{k_E^4} - \cdots \right]
\left[ 1 - \frac{k_E^2}{M^2} + \frac{k_E^4}{M^4} - \cdots \right]\,.
\end{align*}
And hence
\begin{equation*}
\Delta I = 0\,,
\end{equation*}
because each integral has no scale.

Let's stress that the method of regions works thanks to dimensional regularization.
Taylor expansions in each region (i.e., for each set of power counting rules)
must be performed completely, up to the end;
otherwise, integrals in the defect $\Delta I$ can contain some scale(s),
and hence not vanish.

\section{Scalar $\varphi^3$ theory}
\label{S:Scal}

\subsection{Form factor}
\label{S:ScalFF}

\begin{wrapfigure}{r}{0.3\textwidth}
\begin{center}
\vspace{-5mm}
\begin{picture}(40,20)
\put(20,10){\makebox(0,0){\includegraphics{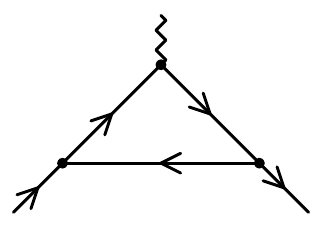}}}
\put(4,2.5){\makebox(0,0){$p$}}
\put(36,2.5){\makebox(0,0){$p'$}}
\put(7,10){\makebox(0,0){$k+p$}}
\put(33,10){\makebox(0,0){$k+p'$}}
\put(20,3){\makebox(0,0){$k$}}
\put(22,17.5){\makebox(0,0){$q$}}
\end{picture}
\end{center}
\caption{Form factor in the scalar $\varphi^3$ theory.}
\label{F:ScalFF}
\end{wrapfigure}
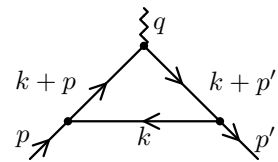
Let's consider massless $\varphi^3$ theory (at $d=4$):
\begin{equation}
L = \frac{1}{2} \left(\partial_\mu\varphi\right) \left(\partial^\mu\varphi\right)
- \frac{g}{3!} \varphi^3\,.
\label{ScalFF:L}
\end{equation}
The dimensionality of the Lagrangian is $[L]=d$,
and hence $[\varphi]=(d-2)/2$, $[g]=(6-d)/2$.
We want to calculate the slightly off-shell form factor
\begin{equation}
I = \int \frac{d^d k}{i\pi^{d/2}}
\frac{1}{[-k^2-i0] [-(k+p)^2-i0] [-(k+p')^2-i0]}
\label{ScalFF:Idef}
\end{equation}
at $-p^2,-p^{\prime2}\ll-q^2$.

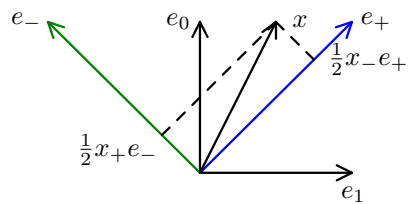
\begin{wrapfigure}{r}{0.4\textwidth}
\begin{center}
\begin{picture}(46,26)
\put(23,13){\makebox(0,0){\includegraphics{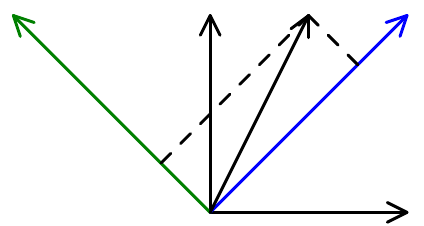}}}
\put(0,23){\makebox(0,0){$e_-$}}
\put(46,23){\makebox(0,0){$e_+$}}
\put(20,23){\makebox(0,0){$e_0$}}
\put(43,0){\makebox(0,0){$e_1$}}
\put(36,23){\makebox(0,0){$x$}}
\put(45,18){\makebox(0,0){$\frac{1}{2} x_- e_+$}}
\put(12,6){\makebox(0,0){$\frac{1}{2} x_+ e_-$}}
\end{picture}
\end{center}
\caption{Light-front components.}
\label{F:lf}
\end{wrapfigure}
It is convenient to use light-front components of vectors.
Instead of the ordinary basis vectors $e_0$, $e_1$ we shall use
\begin{equation*}
e_\pm = e_0 \pm e_1\quad
(e_\pm^2 = 0\,,\quad
e_+ \cdot e_- = 2)\,.
\end{equation*}
The light-front components of a vector $x$ are
\begin{equation}
x_\pm = x \cdot e_\pm = x^0 \mp x^1\,;
\label{ScalFF:lf}
\end{equation}
the vector can be written via them as
\begin{equation*}
x = \frac{1}{2} \left( x_- e_+ + x_+ e_- \right) + x_\bot
\equiv (x_+, x_-, \vec{x}_\bot)\,.
\end{equation*}
The scalar product of two vectors is
\begin{equation*}
x \cdot y = \frac{1}{2} \left( x_+ y_- + x_- y_+ \right)
- \vec{x}_\bot \cdot \vec{y}_\bot\,;
\end{equation*}
in particular, the square of a vector is
\begin{equation*}
x^2 = x_+ x_- - \vec{x}_\bot^{\,2}\,.
\end{equation*}

\begin{wrapfigure}{r}{0.5\textwidth}
\vspace{-10mm}
\begin{center}
\begin{picture}(52,32)
\put(26,16){\makebox(0,0){\includegraphics{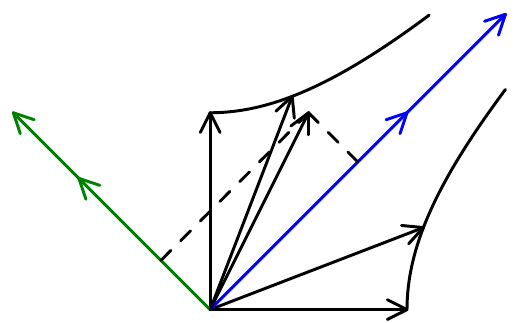}}}
\put(-1,20){\makebox(0,0){$e_-$}}
\put(5.7,13.3){\makebox(0,0){$e'_-$}}
\put(44,20){\makebox(0,0){$e_+$}}
\put(54,30){\makebox(0,0){$e'_+$}}
\put(21,23){\makebox(0,0){$e_0$}}
\put(29.3,24.7){\makebox(0,0){$e'_0$}}
\put(43,1){\makebox(0,0){$e_1$}}
\put(44.7,9.3){\makebox(0,0){$e'_1$}}
\put(33,22){\makebox(0,0){$x$}}
\end{picture}
\end{center}
\caption{A boost.}
\label{F:boost}
\end{wrapfigure}
A boost along the $e_1$ direction
(a Minkowski rotation in the $e_0$, $e_1$ plane)
transforms light-front components in a simple way:
\begin{equation*}
x_\pm' = x_\pm e^{\pm\varphi}\,.
\end{equation*}
The $d$-dimensional volume element is
\begin{equation*}
d^d x = \frac{1}{2} d x_+\,d x_-\,d^{d-2} \vec{x}_\bot\,.
\end{equation*}

The full differential of a function $f(x)$ is
\begin{align*}
d f(x) &{} = \frac{\partial f}{\partial x^0} d x^0 + \frac{\partial f}{\partial x^1} d x^1
+ \frac{\partial f}{\partial\vec{x}_\bot} \cdot d\vec{x}_\bot\\
&{} = \partial^0 f\,dx^0 - \partial^1 f\,dx^1 - \vec{\partial}_\bot f \cdot d\vec{x}_\bot\,,
\end{align*}
and hence
\begin{equation*}
\partial^0 = \frac{\partial}{\partial x^0}\,,\quad
\partial^1 = - \frac{\partial}{\partial x^1}\,,\quad
\vec{\partial}_\bot = - \frac{\partial}{\partial\vec{x}_\bot}\,.
\end{equation*}
In light-front components,
\begin{equation*}
d f(x) = \frac{\partial f}{\partial x_+} dx_+ + \frac{\partial f}{\partial x_-} dx_-
+ \frac{\partial f}{\partial\vec{x}_\bot} \cdot d\vec{x}_\bot
= \frac{1}{2} \partial_+ f\,dx_- + \frac{1}{2} \partial_- f\,dx_+
- \vec{\partial}_\bot \cdot d\vec{x}_\bot\,,
\end{equation*}
and hence
\begin{equation*}
\partial_+ = 2 \frac{\partial}{\partial x_-}\,,\quad
\partial_- = 2 \frac{\partial}{\partial x_+}\,;
\end{equation*}
taking into account
\begin{equation*}
dx_+ = dx^0 - dx^1\,,\quad
dx_- = dx^0 + dx^1\,,
\end{equation*}
we obtain
\begin{equation*}
\partial_+ = \partial^0 - \partial^1\,,\quad
\partial_- = \partial^0 + \partial^1\,,
\end{equation*}
in agreement with~(\ref{ScalFF:lf}).

We choose $e_0$, $e_1$ (and hence $e_+$, $e_-$) in the $p$, $p'$ plane,
then
\begin{align*}
&p = \frac{1}{2} \left( p_- e_+ - \frac{-p^2}{p_-} e_- \right)\,,\quad
p' = \frac{1}{2} \left( - \frac{-p^{\prime2}}{p'_+} e_+ + p'_+ e_- \right)\,,\\
&- q^2 = p_- p'_+ - p^2 - p^{\prime2} + \frac{(-p^2) (-p^{\prime2})}{p_- p'_+}\,;
\end{align*}
our small parameter is
\begin{equation*}
\lambda^2 \sim \frac{-p^2}{-q^2} \sim \frac{-p^{\prime2}}{-q^2} \ll 1\,.
\end{equation*}
In the Breit frame $q^0=0$ (or in any frame related to it by a moderate boost)
light-front components of the external momenta are
\begin{equation}
p \sim (\lambda^2,1,0) Q\,,\quad
p' \sim (1,\lambda^2,0) Q\,,
\label{ScalFF:kin}
\end{equation}
where $Q^2 = -q^2 \approx p_- p'_+$.

\subsection{Method of regions}
\label{S:ScalReg}

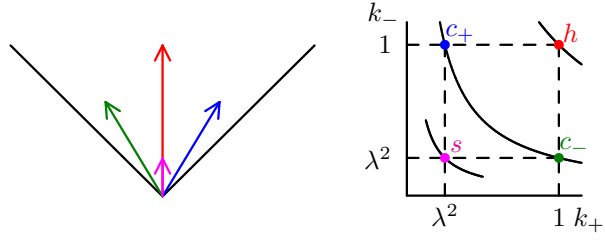
\begin{figure}[h]
\begin{center}
\begin{picture}(78,30)
\put(21,14){\makebox(0,0){\includegraphics{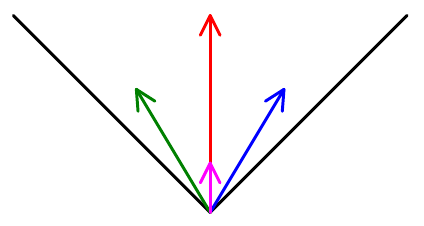}}}
\put(64.5,15.5){\makebox(0,0){\includegraphics{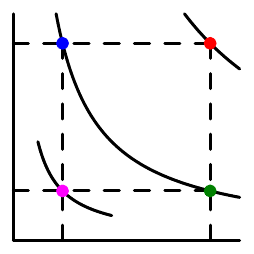}}}
\put(77,0){\makebox(0,0)[b]{$\vphantom{k}\smash{k_+}$}}
\put(73,0){\makebox(0,0)[b]{$1$}}
\put(58,0){\makebox(0,0)[b]{$\lambda^2$}}
\put(52,28){\makebox(0,0)[r]{$k_-$}}
\put(51,24){\makebox(0,0)[r]{$1$}}
\put(51,9){\makebox(0,0)[r]{$\lambda^2$}}
\put(74.5,25.5){\makebox(0,0){$\red{h}$}}
\put(60,25.5){\makebox(0,0){$\blue{c_+}$}}
\put(75,10.5){\makebox(0,0){$\green{c_-}$}}
\put(59.5,10.5){\makebox(0,0){$\magenta{s}$}}
\end{picture}
\end{center}
\caption{Regions in the $k$ space.}
\label{F:ScalReg}
\end{figure}

\begin{wraptable}{r}{0.6\textwidth}
\begin{center}
\vspace{-20mm}
\begin{tabular}{|llll|}
\hline
\textcolor{red}{hard} & $\red{h}$ & $\red{k \sim (1,1,1) Q}$ & $\red{k^2 \sim Q^2}$\\
\textcolor{blue}{collinear} & $\blue{c_+}$ &
$\blue{k \sim (1,\lambda^2,\lambda) Q}$ & $\blue{k^2 \sim Q^2 \lambda^2}$\\
\textcolor{green}{collinear} & $\green{c_-}$ &
$\green{k \sim (\lambda^2,1,\lambda) Q}$ & $\green{k^2 \sim Q^2 \lambda^2}$\\
\textcolor{magenta}{soft} & $\magenta{s}$ &
$\magenta{k \sim (\lambda^2,\lambda^2,\lambda^2) Q}$ & $\magenta{k^2 \sim Q^2 \lambda^4}$\\
\hline
\end{tabular}
\end{center}
\caption{Regions.}
\label{T:Reg}
\end{wraptable}
The form factor~(\ref{ScalFF:Idef}) is given by the sum of 4 regions
(Table~\ref{T:Reg}):
\begin{equation*}
I = \red{I_h} + \blue{I_{c_+}} + \green{I_{c_-}} + \magenta{I_s}\,.
\end{equation*}
Hyperbolas in Fig.~\ref{F:ScalReg} show lines of constant virtuality $k_+ k_-$
(at $\vec{k}_\bot=\vec{0}$).

In the hard region $\red{k \sim (1,1,1) Q}$;
recalling $\blue{p \sim (\lambda^2,1,0) Q}$, $\green{p' \sim (1,\lambda^2,0) Q}$,
we found that the three denominators are
\begin{align*}
& - k^2 = \red{\underbrace{- k_+ k_- + \vec{k}_\bot^{\,2}}_{\mathcal{O}(1)}}\,,\\
& - (k+p)^2 = - (k_+ + p_+) (k_- + p_-) + \vec{k}_\bot^{\,2}
= \red{\underbrace{- k^2 - p_- k_+}_{\mathcal{O}(1)}}
- \underbrace{p_+ (k_- + p_-)}_{\mathcal{O}(\lambda^2)}\,,\\
& - (k+p')^2 = - (k_+ + p'_+) (k_- + p'_-) + \vec{k}_\bot^{\,2}
= \red{\underbrace{- k^2 - p'_+ k_-}_{\mathcal{O}(1)}}
- \underbrace{p'_- (k_+ + p'_+)}_{\mathcal{O}(\lambda^2)}\,,
\end{align*}
and $d^d k \sim \mathcal{O}(1)$.
At the leading order in $\lambda^2$, the hard contribution
\begin{equation*}
\red{I_h} = \int \frac{d^d k}{i\pi^{d/2}}
\frac{1}{\red{(- k^2 - i0) (- k^2 - p_- k_+ - i0) (- k^2 - p'_+ k_- - i0)}}
\end{equation*}
is the form factor with on-shell external legs.
This integral contains a single scale $- q^2 = p_- p'_+$:
if we make the substitution
\begin{equation*}
k_- = p_- \tilde{k}_-\,,\quad
k_+ = p'_+ \tilde{k}_+\,,\quad
\vec{k}_\bot = \sqrt{p_- p'_+} \vec{\tilde{k}}_\bot\,,
\end{equation*}
where $\tilde{k}$ is a dimensionless integration momentum,
then
\begin{equation*}
\red{I_h} = (-q^2)^{d/2-3} \int \frac{d^d \tilde{k}}{i\pi^{d/2}}
\frac{1}{(- \tilde{k}^2 - i0) (- \tilde{k}^2 - \tilde{k}_+ - i0) (- \tilde{k}^2 - \tilde{k}_- - i0)}\,.
\end{equation*}

We shall use Feynman parametrization.
First, we use $\alpha$ parametrization 3 times:
\begin{equation*}
\frac{1}{D_1 D_2 D_3} = \int d\alpha_1 d\alpha_2 d\alpha_3 e^{- \alpha_1 D_1 - \alpha_2 D_2 - \alpha_3 D_3}\,;
\end{equation*}
then we insert $d\eta\,\delta(\alpha_1+\alpha_2+\alpha_3-\eta)$ under the integral sign,
substitute $\alpha_i = \eta x_i$:
\begin{equation*}
\frac{1}{D_1 D_2 D_3} = \int dx_1 dx_2 dx_3 \delta(x_1+x_2+x_3-1) \int d\eta\,\eta^2 e^{-\eta(x_1 D_1 + x_2 D_2 + x_3 D_3)}\,,
\end{equation*}
and integrate in $\eta$:
\begin{equation}
\frac{1}{D_1 D_2 D_3} = \Gamma(3) \int \frac{dx_1 dx_2 dx_3 \delta(x_1+x_2+x_3-1)}{(x_1 D_1 + x_2 D_2 + x_3 D_3)^3}\,.
\label{ScalFF:Feyn}
\end{equation}

We set $-q^2=1$ (its power can be reproduced from dimensions counting),
and use Feynman parametrization:
\begin{equation*}
\red{I_h} = \Gamma(3) \int dx_1\,dx_2\,dx_3\,\delta(x_1+x_2+x_3-1)
\frac{d^d k}{i\pi^{d/2}} \frac{1}{D^3}\,,
\end{equation*}
where the denominator is
\begin{equation*}
D = - x_1 k^2 - x_2 (k^2 + k_+) - x_3 (k^2 + k_-) - i0
= - k^2 - (x_2 e_+ + x_3 e_-) \cdot k - i0\,.
\end{equation*}
Shifting the integration momentum
\begin{equation*}
k' = k + \frac{1}{2} (x_2 e_+ + x_3 e_-)\,,
\end{equation*}
we obtain
\begin{equation*}
D = - k^{\prime2} + x_2 x_3 - i0\,.
\end{equation*}
Now it is easy to calculate the integral in $d^d k'$:
\begin{equation*}
\red{I_h} = \Gamma\left(3-\frac{d}{2}\right) \int_{x_2+x_3<1} dx_2 dx_3 (x_2 x_3)^{d/2-3}\,.
\end{equation*}
Substituting $x_2 = z x$, $x_3 = z (1-x)$, we have
\begin{equation*}
\red{I_h} = \Gamma\left(3-\frac{d}{2}\right) \int_0^1 dz\,z^{d-5} \cdot \int_0^1 dx\,[x(1-x)]^{d/2-3}
= \frac{\Gamma\left(3-\frac{d}{2}\right) \Gamma^2\left(\frac{d}{2}-2\right)}{\Gamma(d-3)}\,.
\end{equation*}
Finally, restoring the power of $-q^2$, we obtain
\begin{equation}
\red{I_h} = \frac{\Gamma(1+\varepsilon)}{\Gamma(1-2\varepsilon)}
\Gamma^2(-\varepsilon) (-q^2)^{-1-\varepsilon}\,.
\label{ScalFF:Ih}
\end{equation}
This integral contains both IR divergence ($k\to0$)
and collinear ones ($k$ non-zero but parallel to $e_+$ or $e_-$);
thus it is $1/\varepsilon^2$.
It contains the single scale $-q^2$.

In the $\blue{c_+}$ collinear region, $\blue{k \sim (\lambda^2,1,\lambda) Q}$;
recalling $\blue{p \sim (\lambda^2,1,0) Q}$, $\green{p' \sim (1,\lambda^2,0) Q}$,
we found that the three denominators are
\begin{align*}
& - k^2 = \blue{\underbrace{- k_+ k_- + \vec{k}_\bot^{\,2}}_{\mathcal{O}(\lambda^2)}}\,,\\
& - (k+p)^2 = \blue{\underbrace{- (k_+ + p_+) (k_- + p_-) + \vec{k}_\bot^{\,2}}_{\mathcal{O}(\lambda^2)}}\,,\\
& - (k+p')^2 = - (k_+ + p'_+) (k_- + p'_-) + \vec{k}_\bot^{\,2}
= \red{\underbrace{- p'_+ k_-}_{\mathcal{O}(1)}}
- \underbrace{(k^2 + p'_+ p'_-)}_{\mathcal{O}(\lambda^2)}
- \underbrace{p'_- k_+}_{\mathcal{O}(\lambda^4)}\,,
\end{align*}
and $d^d k \sim \mathcal{O}(\lambda^d)$.
At the leading order in $\lambda^2$, the $\blue{c_+}$ collinear contribution is
\begin{equation*}
\blue{I_{c_+}} = \frac{1}{p'_+} \int \frac{d^d k}{i\pi^{d/2}} \frac{1}{\blue{(- k^2 - i0)} \blue{(- (k+p)^2 - i0)} \red{(- k_- - i0)}}\,.
\end{equation*}
This integral contains a single scale $- p^2 = - p_+ p_-$:
if we make the substitution
\begin{equation*}
k_+ = (- p_+) \tilde{k}_+\,,\quad
k_- = p_- \tilde{k}_-\,,\quad
\vec{k}_\bot = \sqrt{- p_+ p_-} \vec{\tilde{k}}_\bot\,,
\end{equation*}
where $\tilde{k}$ is a dimensionless integration momentum
and $\tilde{p} = (-1,1,\vec{0})$,
then
\begin{equation*}
\blue{I_{c_+}} = \frac{(- p_+ p_-)^{d/2-2}}{p_- p'_+} \int \frac{d^d\tilde{k}}{i\pi^{d/2}}
\frac{1}{(- \tilde{k}^2 - i0) (- (\tilde{k}+\tilde{p})^2 - i0) (- \tilde{k}_- - i0)}\,.
\end{equation*}

We shall use a variant of Feynman parametrization
in which $d\eta\,\delta(\alpha_1+\alpha_2-\eta)$ is inserted under the integral sign
(the substitutions $\alpha_i = \eta x_i$ are used for all 3 variables $\alpha_i$);
the result is similar to~(\ref{ScalFF:Feyn}), but the $\delta$ function is now $\delta(x_1+x_2-1)$.
The Feynman parameters $x_{1,2}$ correspond to the quadratic denominators;
$x_3$ (which varies from $0$ to $\infty$) -- to the linear one.
Setting the pre-integral factor $(-p^2)^{-\varepsilon}/(-q^2)$ to 1, we obtain
\begin{equation*}
\blue{I_{c_+}} = \Gamma(3) \int dx_1 dx_2 \delta(x_1+x_2-1) dx_3 \frac{d^d k}{i\pi^{d/2}} \frac{1}{D^3}\,,
\end{equation*}
where
\begin{align*}
&D = - x_1 k^2 - x_2 (k+p)^2 - x_3 k_- - i0
= - k^2 - (2 x_2 p + x_3 e_-) \cdot k + x_2 - i0\\
&{} = - k^{\prime2} + x_2 (1 - x_2 + x_3) - i0\,,\qquad
k' = k + x_2 p + \frac{1}{2} x_3 e_-\,,
\end{align*}
and hence
\begin{equation*}
\blue{I_{c_+}} = \Gamma\left(3 - \frac{d}{2}\right) \int dx_1 dx_3 [x_1 (1 - x_1 + x_3)]^{d/2 - 3}\,.
\end{equation*}
Substituting $x_3 = (1-x_1) z$ we have
\begin{align*}
&\blue{I_{c_+}} = \Gamma\left(3 - \frac{d}{2}\right) \int_0^\infty dz\,(1+z)^{d/2-3} \cdot \int_0^1 dx_1\,x_1^{d/2-3} (1-x_1)^{d/2-2}\\
&{} = \frac{\Gamma\left(2-\frac{d}{2}\right) \Gamma\left(\frac{d}{2}-2\right) \Gamma\left(\frac{d}{2}-1\right)}{\Gamma(d-3)}\,.
\end{align*}
Finally, restoring the pre-integral factor, we arrive at
\begin{equation}
\blue{I_{c_+}} = \frac{\Gamma(1-\varepsilon)}{\Gamma(1-2\varepsilon)}
\Gamma(\varepsilon) \Gamma(-\varepsilon) \frac{(-p^2)^{-\varepsilon}}{-q^2}\,.
\label{ScalFF:Ic}
\end{equation}
Of course,
\begin{equation*}
\green{I_{c_-}} = \frac{\Gamma(1-\varepsilon)}{\Gamma(1-2\varepsilon)}
\Gamma(\varepsilon) \Gamma(-\varepsilon) \frac{(-p^{\prime2})^{-\varepsilon}}{-q^2}\,.
\end{equation*}

In the soft region $\magenta{k \sim (\lambda^2,\lambda^2,\lambda^2) Q}$;
recalling $\blue{p \sim (\lambda^2,1,0) Q}$, $\green{p' \sim (1,\lambda^2,0) Q}$,
we found that the three denominators are
\begin{align*}
& - k^2 = \magenta{\underbrace{- k_+ k_- + \vec{k}_\bot^{\,2}}_{\mathcal{O}(\lambda^4)}}\,,\\
& - (k+p)^2 = - (k_+ + p_+) (k_- + p_-) + \vec{k}_\bot^{\,2}
= \blue{\underbrace{- p_- (k_+ + p_+)}_{\mathcal{O}(\lambda^2)}}
- \underbrace{(k^2 + p_+ k_-)}_{\mathcal{O}(\lambda^4)}\,,\\
& - (k+p')^2 = - (k_+ + p'_+) (k_- + p'_-) + \vec{k}_\bot^{\,2}
= \green{\underbrace{- p'_+ (k_- + p'_-)}_{\mathcal{O}(\lambda^2)}}
- \underbrace{(k^2 + p'_- k_+)}_{\mathcal{O}(\lambda^4)}\,,
\end{align*}
and $d^d k \sim \mathcal{O}(\lambda^{2d})$.
At the leading order in $\lambda^2$, the soft contribution is
\begin{equation*}
\magenta{I_s} = \frac{1}{p_- p'_+} \int \frac{d^d k}{i\pi^{d/2}}
\frac{1}{\magenta{(- k^2 - i0)} \blue{(- k_+ - p_+ - i0)} \green{(- k_- - p'_- - i0)}}\,.
\end{equation*}
This integral contains a single scale
\begin{equation*}
\frac{(-p^2) (-p^{\prime2})}{-q^2} = \underbrace{p_+ p'_-}_{\mathcal{O}(\lambda^4)}
+ \mathcal{O}(\lambda^6)\,:
\end{equation*}
if we make the substitution
\begin{equation*}
k_+ = (- p_+) \tilde{k}_+\,,\quad
k_- = (- p'_-) \tilde{k}_-\,,\quad
\vec{k}_\bot = \sqrt{(- p_+) (- p'_-)} \vec{\tilde{k}}_\bot\,,
\end{equation*}
where $\tilde{k}$ is a dimensionless integration momentum, then
\begin{equation*}
\magenta{I_s} = \frac{(p_+ p'_-)^{d/2-2}}{p_- p'_+} \int \frac{d^d\tilde{k}}{i\pi^{d/2}}
\frac{1}{(- \tilde{k}^2 - i0) (- \tilde{k}_+ + 1 - i0) (- \tilde{k}_- + 1 - i0)}\,.
\end{equation*}
This integral is similar to off-shell HQET ones (but with light-like directions $e_+$, $e_-$);
it is UV divergent.

We shall use a variant of Feynman parametrization~(\ref{ScalFF:Feyn})
with $\delta(x_1-1)$ where $x_1$ corresponds to the quadratic denominator.
Setting the scale factor $[(-p^2)(-p^{\prime2})]^{-\varepsilon}/(-q^2)^{1-\varepsilon}$ to 1,
we obtain
\begin{equation*}
\magenta{I_s} = \Gamma(3) \int dx_2 dx_3 \frac{d^d k}{i\pi^{d/2}} \frac{1}{D^3}\,,
\end{equation*}
where
\begin{align*}
&D = - k^2 + x_2 (- k_+ + 1) + x_3 (- k_- + 1) - i0
= - k^2 - (x_2 e_+ + x_3 e_-) \cdot k + x_2 + x_3 - i0\\
&{} = - k^{\prime2} + x_2 + x_3 + x_2 x_3 - i0\,,\qquad
k' = k + \frac{1}{2} (x_2 e_+ + x_3 e_-)\,,
\end{align*}
and hence
\begin{equation*}
\magenta{I_s} = \Gamma\left(3 - \frac{d}{2}\right)
\int dx_2 dx_3 (x_2 + x_3 + x_2 x_3)^{d/2 - 3}\,.
\end{equation*}
Substituting $x_2 = y x$, $x_3 = y (1-x)$, we have
\begin{equation*}
\magenta{I_s} = \Gamma\left(3 - \frac{d}{2}\right)
\int_0^\infty dy\,y^{d/2-2}\,\int_0^1 dx\,[1 + y x (1-x)]^{d/2-3}\,.
\end{equation*}
Substituting $y x (1-x) = z$, we have
\begin{equation*}
\magenta{I_s} = \Gamma\left(3 - \frac{d}{2}\right)
\int_0^\infty dz\,z^{d/2-2} (1+z)^{d/2-3} \cdot
\int_0^1 dx\,[x(1-x)]^{1-d/2}\,.
\end{equation*}
Substituting $1+z = 1/u$, we obtain
\begin{equation*}
\magenta{I_s} = \Gamma\left(3 - \frac{d}{2}\right)
\int_0^1 du\,u^{3-d} (1-u)^{d/2-2} \cdot
\int_0^1 dx\,[x(1-x)]^{1-d/2}
= \Gamma\left(\frac{d}{2}-1\right) \Gamma^2\left(2-\frac{d}{2}\right)\,.
\end{equation*}
Finally, restoring the pre-integral factor, we arrive at
\begin{equation}
\magenta{I_s} = \Gamma(1-\varepsilon) \Gamma^2(\varepsilon)
\frac{\left[(-p^2) (-p^{\prime2})\right]^{-\varepsilon}}{(-q^2)^{1-\varepsilon}}\,.
\label{ScalFF:Is}
\end{equation}

Now we can combine $\red{I_h}$~(\ref{ScalFF:Ih}),
$\blue{I_{c_+}}$~(\ref{ScalFF:Ic}), $\green{I_{c_-}}$,
and $\magenta{I_s}$~(\ref{ScalFF:Is}):
\begin{align*}
I ={}& \frac{\Gamma(1+\varepsilon)}{\Gamma(1-2\varepsilon)}
\Gamma^2(-\varepsilon) (-q^2)^{-1-\varepsilon}\\
&{}\times \left[ 1
- \left(\frac{-p^2}{-q^2}\right)^{-\varepsilon}
- \left(\frac{-p^{\prime2}}{-q^2}\right)^{-\varepsilon}
+ \frac{\Gamma(1+\varepsilon) \Gamma(1-2\varepsilon)}{\Gamma(1-\varepsilon)}
\left(\frac{-p^2}{-q^2}\right)^{-\varepsilon}
\left(\frac{-p^{\prime2}}{-q^2}\right)^{-\varepsilon}
\right]\,.
\end{align*}
The diagram~(\ref{ScalFF:Idef}) we are calculating is finite at $\varepsilon\to0$;
and indeed, all divergences cancel, and we obtain
\begin{equation}
I = \frac{1}{-q^2}
\left( \log\frac{-p^2}{-q^2}\,\log\frac{-p^{\prime2}}{-q^2} + \frac{\pi^2}{3} \right)
+ \mathcal{O}(\lambda^2)\,.
\label{ScalFF:I}
\end{equation}
This cancellation of $1/\varepsilon^2$ and $1/\varepsilon$ divergences
is a good check that we have not forgotten some region.

What does happen if we invent some other region and calculate its contribution?
For example, consider a region $\brown{k \sim (\lambda,\lambda,\lambda) Q}$
(let's call it semihard).
The denominators are
\begin{align*}
& - k^2 = \brown{\underbrace{- k_+ k_- + \vec{k}_\bot^{\,2}}_{\mathcal{O}(\lambda^2)}}\,,\\
& - (k+p)^2 = - (k_+ + p_+) (k_- + p_-) + \vec{k}_\bot^{\,2}
= \brown{\underbrace{- p_- k_+}_{\mathcal{O}(\lambda)}} + \mathcal{O}(\lambda^2)\,,\\
& - (k+p')^2 = - (k_+ + p'_+) (k_- + p'_-) + \vec{k}_\bot^{\,2}
= \brown{\underbrace{- p'_+ k_-}_{\mathcal{O}(\lambda)}} + \mathcal{O}(\lambda^2)\,.
\end{align*}
The contribution of this region is
\begin{equation*}
\brown{I_{sh}} = \frac{1}{p_- p'_+} \int \frac{d^d k}{i\pi^{d/2}}
\frac{1}{(- k^2 - i0) (- k_+ - i0) (- k_- - i0)} = 0
\end{equation*}
because the integral contains no scale.
Similarly, with some work, one can check that there exist no regions
of the form $k \sim (\lambda^a,\lambda^b,\lambda^c) Q$ producing non-zero contributions to $I$
except the ones we have considered in this section.
This is not a complete proof that there are no extra regions
if they don't have such a form in terms of the light-front components,
but their existence is highly unlikely.

\subsection{SCET Lagrangian}
\label{S:SCET}

Instead of using the method of regions for diagrams in the full theory,
we can formulate a low-energy effective field theory.
Hard contributions are integrated out;
they only appear in matching coefficients which accumulate information about physics at small distances.
Processes at larger distances,
namely, collinear and soft contributions (Table~\ref{T:Reg}, Fig.~\ref{F:ScalReg}),
are explicitly considered in the effective theory.
Instead of a single field $\varphi$ in the full theory,
we now have 3 fields:
\begin{equation*}
\varphi \to \blue{\varphi_{c_+}} + \green{\varphi_{c_-}} + \magenta{\varphi_s}\,.
\end{equation*}
Substituting this field decomposition into the full theory Lagrangian~(\ref{ScalFF:L}),
we obtain
\begin{equation*}
L = \blue{L_{c_+}} + \green{L_{c_-}} + \magenta{L_s} + L_{cs}\,,
\end{equation*}
where
\begin{align*}
&\blue{L_{c_+} = \frac{1}{2} (\partial_\mu \varphi_{c_+}) (\partial^\mu \varphi_{c_+})
- \frac{g}{3!} \varphi_{c_+}^3}\,,\\
&\green{L_{c_-} = \frac{1}{2} (\partial_\mu \varphi_{c_-}) (\partial^\mu \varphi_{c_-})
- \frac{g}{3!} \varphi_{c_-}^3}\,,\\
&\magenta{L_s = \frac{1}{2} (\partial_\mu \varphi_s) (\partial^\mu \varphi_s)
- \frac{g}{3!} \varphi_s^3}
\end{align*}
are 3 copies of the Lagrangian~(\ref{ScalFF:L}),
and $L_{cs}$ describes interactions of collinear and soft fields.

Our small parameter is $\lambda$;
each quantity in the effective theory scales as some power of $\lambda$.
Soft field momenta scale as $\magenta{k_s \sim \lambda^2}$;
this means $\magenta{\partial_s \sim \lambda^2}$ where $\magenta{\partial_s}$ acts on soft fields,
and $\magenta{x_s \sim \lambda^{-2}}$ where $\magenta{x_s}$ is a distance at which soft fields substantially change.
In order to find power counting for the soft field $\magenta{\varphi_s}$,
we consider the correlator
\begin{equation*}
\magenta{{<}T \varphi_s(x_s) \varphi_s(0){>} = \int \frac{d^d k_s}{(2\pi)^d} e^{-i k_s\cdot x_s} \frac{i}{k_s^2+i0} \sim \lambda^{2d-4}}\,,
\end{equation*}
and hence
\begin{equation*}
\magenta{\varphi_s \sim \lambda^{d-2}}\,.
\end{equation*}
Therefore, the Lagrangian is
\begin{equation*}
\magenta{L_s \sim (\partial_\mu \varphi_s) (\partial^\mu \varphi_s) \sim \lambda^{2d}}\,,
\end{equation*}
and the action for a region of size $\magenta{\sim x_s}$ is
\begin{equation*}
\magenta{S_s = \int d^d x\,L_s \sim 1}\,.
\end{equation*}
This is natural:
the characteristic action is $\sim1$, not a small correction,
and we cannot expand in it when calculating the path integral.

\begin{sloppypar}
For $\blue{c_+}$ collinear fields $\blue{k_{c_+} \sim (\lambda^2,1,\lambda)}$
and $\blue{\partial_{c_+} \sim (\lambda^2,1,\lambda)}$;
from $\blue{k_+ x_- \sim 1}$, $\blue{k_- x_+ \sim 1}$, $\blue{k_\bot x_\bot \sim 1}$
we find $\blue{x_{c_+} \sim (1,\lambda^{-2},\lambda^{-1})}$.
The correlator of collinear fields is
\begin{equation*}
\blue{{<}T \varphi_{c_+}(x) \varphi_{c_+}(0){>} =
\int \frac{d^d k_{c_+}}{(2\pi)^d} e^{-i k_{c_+}\cdot x_{c_+}} \frac{i}{k_{c_+}^2+i0} \sim \lambda^{d-2}}\,,
\end{equation*}
and hence
\begin{equation*}
\blue{\varphi_{c_+} \sim \lambda^{d/2-1}}\,.
\end{equation*}
Therefore, the Lagrangian is
\begin{equation*}
\blue{L_{c_+} \sim (\partial_+ \varphi_{c_+}) (\partial_- \varphi_{c_+}) \sim \lambda^d}\,,
\end{equation*}
and the action for a characteristic region $\blue{\sim x_{c_+}}$ is
\begin{equation*}
\blue{S_{c_+} = \int d^d x\,L_{c_+} \sim 1}\,.
\end{equation*}
Similar estimates can be made for $\green{\varphi_{c_-}}$.
\end{sloppypar}

\begin{figure}[h]
\begin{center}
\begin{picture}(74,20)
\put(10,11.5){\makebox(0,0){\includegraphics{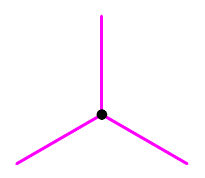}}}
\put(10,0){\makebox(0,0)[b]{a}}
\put(36,11.5){\makebox(0,0){\includegraphics{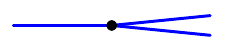}}}
\put(36,0){\makebox(0,0)[b]{b}}
\put(63,11.5){\makebox(0,0){\includegraphics{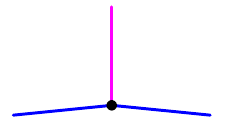}}}
\put(63,0){\makebox(0,0)[b]{c}}
\end{picture}
\end{center}
\caption{Interactions: 3 soft fields (a); 3 $\blue{c_+}$ collinear fields (b);
2 $\blue{c_+}$ collinear fields and soft field (c).}
\label{F:Int}
\end{figure}
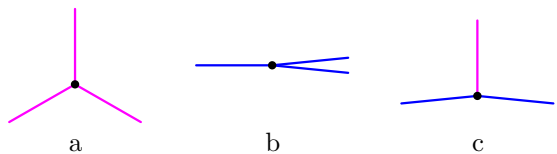

There is interaction of 3 soft fields (Fig.~\ref{F:Int}a)
from the term $\magenta{- \frac{g}{3!} \varphi_s^3}$ in $\magenta{L_s}$;
momentum conservation is consistent with power counting:
$\magenta{(\lambda^2,\lambda^2,\lambda^2)+(\lambda^2,\lambda^2,\lambda^2)=(\lambda^2,\lambda^2,\lambda^2)}$.
There is interaction of 3 $\blue{c_+}$ collinear fields (Fig.~\ref{F:Int}b)
from the term $\blue{ - \frac{g}{3!} \varphi_{c_+}^3}$ in $\blue{L_{c_+}}$;
momentum conservation is consistent:
$\blue{(\lambda^2,1,\lambda)+(\lambda^2,1,\lambda)=(\lambda^2,1,\lambda)}$
(of course, there is also interaction of 3 $\green{c_-}$ collinear fields).
There is also interaction of 2 $\blue{c_+}$ collinear fields and soft one (Fig.~\ref{F:Int}c)
from the collinear--soft Lagrangian
\begin{equation}
L_{cs} = - \frac{g}{2} \blue{\varphi_{c_+}^2} \magenta{\varphi_s} - \frac{g}{2} \green{\varphi_{c_-}^2} \magenta{\varphi_s}
\label{SCET:Lcs}
\end{equation}
momentum conservation is consistent:
$\blue{(\lambda^2,1,\lambda)}+\magenta{(\lambda^2,\lambda^2,\lambda^2)}=\blue{(\lambda^2,1,\lambda)}$
(of course, there is also interaction of 2 $\green{c_-}$ collinear fields and soft one).
Other possible interactions
($\blue{\varphi_{c_+}}^2 \green{\varphi_{c_-}}$,
$\blue{\varphi_{c_+}} \green{\varphi_{c_-}} \magenta{\varphi_s}$,
$\blue{\varphi_{c_+}}^2 \magenta{\varphi_s}$, \dots)
are not allowed because momentum conservation is not consistent with power counting.

\begin{wrapfigure}[6]{r}{0.3\textwidth}
\begin{center}
\vspace{-5mm}
\includegraphics{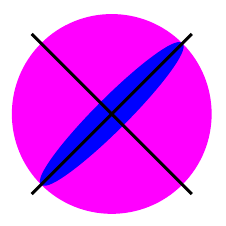}
\end{center}
\caption{Collinear--soft interaction in coordinate space.}
\label{F:Intcs}
\end{wrapfigure}

The action for the collinear--soft interaction is
\begin{equation}
S_{cs} = \frac{g}{2} \int d^d x\,\blue{\varphi_{c_+}^2(x)} \magenta{\varphi_s(x)}
\label{SCET:Scs}
\end{equation}
(the $\green{c_-}$--soft interaction can be considered similarly).
In coordinate space the region where $\blue{\varphi_{c_+}(x)}$ lives
is $\blue{x \sim (1,\lambda^{-2},\lambda^{-1}) Q^{-1}}$.
The region where $\magenta{\varphi_s(x)}$ lives
is $\magenta{x \sim (\lambda^{-2},\lambda^{-2},\lambda^{-2}) Q^{-1}}$.
The characteristic region of $x$ in the integral~(\ref{SCET:Scs})
is therefore collinear (Fig.~\ref{F:Intcs}).

\noindent
The soft field $\magenta{\varphi_s(\blue{x})}$ varies little in the $e_-$ direction
within this region:
$\magenta{\varphi_s(\blue{x}) \approx \varphi_s(\blue{\bar{x}_+})}$,
where $\bar{x}_+ \equiv \frac{1}{2} x_- e_+^\mu$ is the projection of $x$ onto the $e_+$ axis.
Derivatives acting on $\magenta{\varphi_s(x)}$
are $\magenta{\partial_s \sim (\lambda^2,\lambda^2,\lambda^2) Q}$,
and we can expand
\begin{align*}
\magenta{\varphi_s(\blue{x})} ={}& \magenta{\varphi_s(\blue{\bar{x}_+})}
- \underbrace{\blue{\vec{x}_\bot} \cdot \magenta{\vec{\partial}_\bot}}_{\mathcal{O}(\lambda)} \magenta{\varphi_s(\blue{\bar{x}_+})}
+ \frac{1}{2} \underbrace{\blue{x_+} \magenta{\partial_-}}_{\mathcal{O}(\lambda^2)} \magenta{\varphi_s(\blue{\bar{x}_+})}\\
&{} + \frac{1}{2} \underbrace{\blue{x_{\bot i} x_{\bot j}} \magenta{\partial_{\bot i} \partial_{\bot j}}}_{\mathcal{O}(\lambda^2)} \magenta{\varphi_s(\blue{\bar{x}_+})}
+ \mathcal{O}(\lambda^3)\,.
\end{align*}
This is called multipole expansion, in analogy with electrodynamics.
Therefore, the action~(\ref{SCET:Scs}) is
\begin{equation*}
S_{cs} = \frac{g}{2} \int d^d \blue{x}\,\blue{\varphi_{c_+}^2(x)} \magenta{\varphi_s(\blue{\bar{x}_+})}
+ \mathcal{O}(\lambda)\,.
\end{equation*}

We can also understand this result from a slightly different point of view.
Expressing the coordinate-space fields in the action~(\ref{SCET:Scs}) via the momentum-space ones
we obtain
\begin{equation*}
S_{cs} = \frac{g}{2} \int d^d x
\frac{d^d\blue{k_{c_+1}}}{(2\pi)^d} \frac{d^d\blue{k_{c_+2}}}{(2\pi)^d} \frac{d^d\magenta{k_s}}{(2\pi)^d}
e^{-i (\blue{k_{c_+1}} + \blue{k_{c_+2}} + \magenta{k_s}) \cdot x}
\blue{\tilde{\varphi}_{c_+}(k_{c_+1})} \blue{\tilde{\varphi}_{c_+}(k_{c_+2})} \magenta{\tilde{\varphi}_s(k_s)}\,.
\end{equation*}
Here $\blue{k_{c_+1}} + \blue{k_{c_+2}} + \magenta{k_s} \sim \blue{(\lambda^2,1,\lambda) Q}$,
and hence the characteristic $\blue{x \sim (1,\lambda^{-2},\lambda^{-1}) Q^{-1}}$;
for a soft momentum $\magenta{k_s \sim (\lambda^2,\lambda^2,\lambda^2) Q}$ we have
\begin{equation*}
\magenta{k_s} \cdot \blue{x}
= \frac{1}{2} \underbrace{\magenta{k_{s+}} \blue{x_-}}_{\mathcal{O}(1)}
+ \frac{1}{2} \underbrace{\magenta{k_{s-}} \blue{x_+}}_{\mathcal{O}(\lambda^2)}
- \underbrace{\magenta{\vec{k}_{s\bot}} \cdot \blue{\vec{x}_\bot}}_{\mathcal{O}(\lambda)}\,.
\end{equation*}
In other words, the soft field $\magenta{\varphi_s}$,
when interacting with the collinear field $\blue{\varphi_{c_+}}$,
effectively carries momentum $\magenta{k_s=(k_{s+},0,\vec{0})}$.
The integral in $d^d\magenta{k_s}$ becomes
\begin{equation*}
\int \frac{d^d\magenta{k_s}}{(2\pi)^d} e^{i \magenta{k_{s+}} x_-} \magenta{\tilde{\varphi}_s(k_s)}
= \magenta{\varphi_s(\bar{x}_+)}\,.
\end{equation*}

Now we can write down the complete SCET Lagrangian:
\begin{align}
L ={}& \blue{\frac{1}{2} (\partial_\mu \varphi_{c_+}(x)) (\partial^\mu \varphi_{c_+}(x))}
- \blue{\frac{g}{3!} \varphi_{c_+}^3(x)}
\nonumber\\
&{} + \green{\frac{1}{2} (\partial_\mu \varphi_{c_-}(x)) (\partial^\mu \varphi_{c_-}(x))}
- \green{\frac{g}{3!} \varphi_{c_-}^3(x)}
\nonumber\\
&{} + \magenta{\frac{1}{2} (\partial_\mu \varphi_s(x)) (\partial^\mu \varphi_s(x))}
- \magenta{\frac{g}{3!} \varphi_s^3(x)}
\nonumber\\
&{} - \frac{g}{2} \blue{\varphi_{c_+}^2(x)} \magenta{\varphi_s(\bar{x}_+)}
- \frac{g}{2} \green{\varphi_{c_-}^2(x)} \magenta{\varphi_s(\bar{x}_-)}
+ \mathcal{O}(\lambda)\,.
\label{SCET:L}
\end{align}
It is not exactly translation invariant;
however, it is translation invariant up to $\mathcal{O}(\lambda)$.
It is possible to include $\mathcal{O}(\lambda)$, $\mathcal{O}(\lambda^2)$, etc.,
correction terms to this Lagrangian;
then it will become translation invariant up to higher powers of $\lambda$.

As discussed in Sect.~\ref{S:Reg}, all quantities should be \emph{completely} expanded in $\lambda$,
i.\,e., each term should be proportional to some power $\lambda^n$,
not a nontrivial function with all powers of $\lambda$;
otherwise, the argument about the absence of double counting in dimensional regularization will not work.
Therefore we should use not the local Lagrangian~(\ref{SCET:Lcs}) (which contains all powers of $\lambda$)
but the multipole-expanded nonlocal Lagrangian~(\ref{SCET:L}).

Generally speaking, coefficients in effective Lagrangians are obtained by matching,
and can contain radiative corrections (from hard loops).
But there are no loop corrections to coefficients in the SCET Lagrangian~(\ref{SCET:L}).
For example, let's consider the $\blue{c_+}$ collinear interaction,
and suppose that it contains an unknown matching coefficient:
\begin{equation*}
\blue{L_{c_+} = - \frac{g}{3!} C \varphi_{c_+}^3\,,\quad
C = 1 + C_1 g^2 + \cdots}
\end{equation*}
We calculate the scattering amplitude with all 3 external momenta $\blue{c_+}$ collinear
both in the full theory and in the effective one.
With one-loop accuracy
\begin{equation*}
\raisebox{-3mm}{\includegraphics{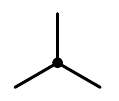}}
+ \raisebox{-5mm}{\includegraphics{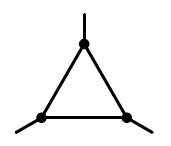}}
= C \raisebox{-3mm}{\includegraphics{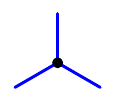}}
+ \raisebox{-5mm}{\includegraphics{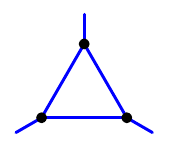}}
+ \raisebox{-5mm}{\includegraphics{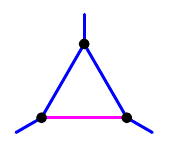}}
+ \raisebox{-5mm}{\includegraphics{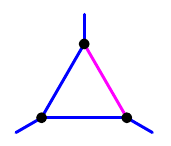}}
+ \raisebox{-5mm}{\includegraphics{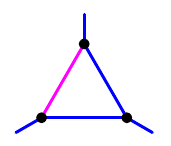}}\,.
\end{equation*}
When all 3 external momenta are exactly parallel to $e_+$ (and hence on-shell),
all loop diagrams vanish, and we obtain
\begin{equation*}
\blue{C} = 1\,.
\end{equation*}

\begin{wrapfigure}{r}{0.4\textwidth}
\begin{center}
\vspace{-5mm}
\begin{picture}(50,22)
\put(11,11){\makebox(0,0){\includegraphics{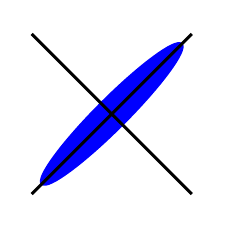}}}
\put(39,11){\makebox(0,0){\includegraphics{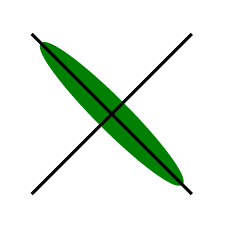}}}
\end{picture}
\end{center}
\caption{Coordinate-space regions where $\blue{c_+}$ and $\green{c_-}$ collinear fields live.}
\label{F:xcoll}
\end{wrapfigure}
Let's now consider the current $J(x) = \frac{1}{2} \varphi^2(x)$ in the full theory
in the kinematical situation when it annihilates one (or more) $\blue{c_+}$ collinear particle(s)
and creates one (or more) $\green{c_-}$ collinear particle(s).
In SCET it becomes a sum of 2-field, 3-field (and so on) operators:
$J(x) = J_2(x) + J_3(x) + \cdots$,
where symbolically
\begin{equation*}
J_2 = \red{C_2} \blue{\varphi_{c_+}} \green{\varphi_{c_-}}\,,\quad
J_3 = \frac{1}{2} \red{C_3} \left[ \blue{\varphi_{c_+}^2} \green{\varphi_{c_-}} + \blue{\varphi_{c_+}} \green{\varphi_{c_-}^2} \right]\,,
\,\ldots
\end{equation*}
The field $\blue{\varphi_{c_+}(x)}$ lives in the region $\blue{x \sim (1,\lambda^{-2},\lambda^{-1}) Q^{-1}}$
while $\green{\varphi_{c_-}(x)}$ -- in $\green{x \sim (\lambda^{-2},1,\lambda^{-1}) Q^{-1}}$ (Fig.~\ref{F:xcoll}).
The field $\blue{\varphi_{c_+}(x)}$ varies quickly in the $e_-$ direction:
for $t\sim1/Q$
\begin{equation*}
\blue{\varphi_{c_+}}(x + t e_-) = \blue{\varphi_{c_+}}(x)
+ \underbrace{t \partial_-}_{\mathcal{O}(1)} \blue{\varphi_{c_+}}(x)
+ \frac{1}{2} \underbrace{t^2 \partial_-^2}_{\mathcal{O}(1)} \blue{\varphi_{c_+}}(x)
+ \cdots\,,
\end{equation*}
cannot be expanded in $t$.
Similarly,
\begin{equation*}
\green{\varphi_{c_-}}(x + t' e_+) = \green{\varphi_{c_-}}(x)
+ \underbrace{t' \partial_+}_{\mathcal{O}(1)} \green{\varphi_{c_-}}(x)
+ \frac{1}{2} \underbrace{t^{\prime2} \partial_+^2}_{\mathcal{O}(1)} \green{\varphi_{c_-}}(x)
+ \cdots
\end{equation*}
cannot be expanded in $t'\sim1/Q$.
It is natural to expect nonlocality of order $Q^{-1}$ in $J_2$ (as well as $J_3$, etc.).
We can expand $\blue{\varphi_{c_+}(x)}$ in shifts in all directions except $e_-$,
and $\green{\varphi_{c_-}(x)}$ -- in all directions except $e_+$:
\begin{equation*}
J_2(x) = \int d\blue{t}\,d\green{t'}\,\red{C_2}(\blue{t},\green{t'}) \blue{\varphi_{c_+}(x + t e_-)} \green{\varphi_{c_-}(x + t' e_+)}
\end{equation*}
plus terms suppressed by powers of $\lambda$.

In momentum space, for on-shell momenta $\blue{p = (0,p_-,\vec{0})}$ and $\green{p' = (p'_+,0,\vec{0})}$
we have the matching condition
\begin{equation*}
\raisebox{-5mm}{\begin{picture}(14,11)
\put(7,5.5){\makebox(0,0){\includegraphics{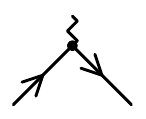}}}
\put(1,4){\makebox(0,0){$\blue{p}$}}
\put(13,4){\makebox(0,0){$\green{p'}$}}
\end{picture}}
= \red{\tilde{C}_2}(\blue{p_-},\green{p'_+})
\raisebox{-5mm}{\begin{picture}(14,11)\put(7,5.5){\makebox(0,0){\includegraphics{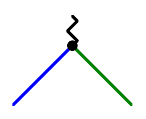}}}\end{picture}}\,,\quad
\red{\tilde{C}_2}(\blue{p_-},\green{p'_+}) = \int d\blue{t}\,d\green{t'}\,e^{i \blue{p_- t} - i \green{p'_+ t'}} C_2(\blue{t},\green{t'})\,.
\end{equation*}
At the tree level there is no nonlocality:
\begin{equation*}
\red{\tilde{C}_2^{(0)}}(\blue{p_-},\green{p'_+}) = 1\,,\quad
\red{C_2^{(0)}}(\blue{t},\green{t'}) = \delta(\blue{t}) \delta(\green{t'})\,.
\end{equation*}
With 1-loop accuracy,
\begin{equation*}
\raisebox{-5mm}{\begin{picture}(24,14)
\put(12,7){\makebox(0,0){\includegraphics{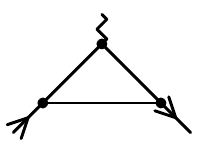}}}
\put(2,3){\makebox(0,0){$\blue{p}$}}
\put(22,3){\makebox(0,0){$\green{p'}$}}
\end{picture}}
= \red{\tilde{C}_2^{(1)}}(\blue{p_-},\green{p'_+})
\raisebox{-3mm}{\includegraphics{grozin_andrey_fig23.pdf}}
+ \raisebox{-5mm}{\includegraphics{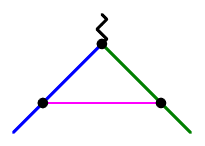}}
\end{equation*}
The SCET loop vanishes at on-shell momenta $\blue{p = (0,p_-,\vec{0})}$, $\green{p' = (p'_+,0,\vec{0})}$,
and the 1-loop correction to $\tilde{C}_2(\blue{p_-},\green{p'_+})$
is given by the hard loop $\red{I_h}$~(\ref{ScalFF:Ih}):
\begin{equation}
\red{\tilde{C}_2^{(1)}}(\blue{p_-},\green{p'_+}) = \frac{g^2}{(4\pi)^{d/2}} \red{I_h}(\blue{p_-} \green{p'_+})
\label{SCET:C2}
\end{equation}
(note that it depends only on the product $\blue{p_-} \green{p'_+}$).

Similarly, the SCET operator $J_3$ is
\begin{equation*}
J_3(x) = \frac{1}{2} \int d\blue{t_1}\,d\blue{t_2}\,d\green{t'}\,\red{C_3}(t_1,t_2,t')
\blue{\varphi_{c_+}(x + t_1 e_-)} \blue{\varphi_{c_+}(x + t_2 e_-)} \green{\varphi_{c_-}(x + t' e_+)}
+ ({+}\leftrightarrow{-})\,.
\end{equation*}
The leading-order matching condition is
\begin{align*}
&\raisebox{-10mm}{\begin{picture}(32,22)
\put(16,11){\makebox(0,0){\includegraphics{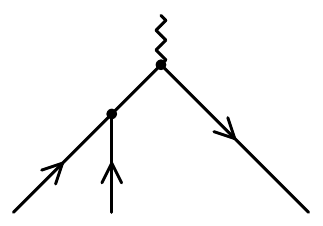}}}
\put(0,3){\makebox(0,0){$\blue{p_1}$}}
\put(13,3){\makebox(0,0){$\blue{p_2}$}}
\put(32,3){\makebox(0,0){$\green{p'}$}}
\end{picture}}
+ \raisebox{-10mm}{\begin{picture}(32,22)
\put(16,11){\makebox(0,0){\includegraphics{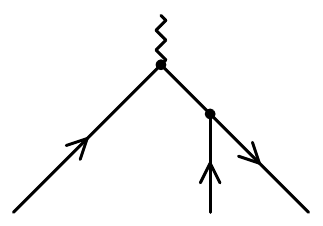}}}
\put(0,3){\makebox(0,0){$\blue{p_1}$}}
\put(19,3){\makebox(0,0){$\blue{p_2}$}}
\put(32,3){\makebox(0,0){$\green{p'}$}}
\end{picture}}
+ (1\leftrightarrow2)\\
&{} = \raisebox{-10mm}{\begin{picture}(32,22)
\put(16,11){\makebox(0,0){\includegraphics{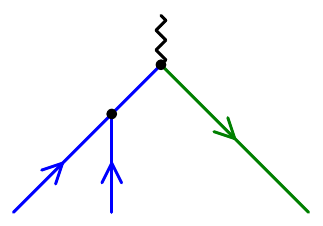}}}
\put(0,3){\makebox(0,0){$\blue{p_1}$}}
\put(13,3){\makebox(0,0){$\blue{p_2}$}}
\put(32,3){\makebox(0,0){$\green{p'}$}}
\end{picture}}
+ \red{\tilde{C}_3}
\raisebox{-10mm}{\begin{picture}(32,22)
\put(16,11){\makebox(0,0){\includegraphics{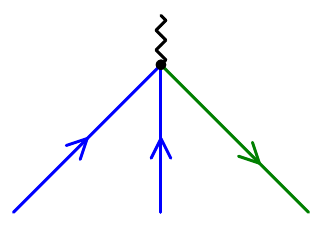}}}
\put(0,3){\makebox(0,0){$\blue{p_1}$}}
\put(14,3){\makebox(0,0){$\blue{p_2}$}}
\put(32,3){\makebox(0,0){$\green{p'}$}}
\end{picture}}
\end{align*}
The first diagram on the second line is identical to the first diagram on the first line,
and we may cancel them.
After that, we may put the momenta on-shell,
$\blue{p_1 = (0,p_{1-},\vec{0})}$, $\blue{p_2 = (0,p_{2-},\vec{0})}$, $\green{p' = (p'_+,0,\vec{0})}$,
and obtain
\begin{equation*}
\red{\tilde{C}_3}(\blue{p_{1-}},\blue{p_{2-}},\green{p'_+}) =
\raisebox{-10mm}{\begin{picture}(32,22)
\put(16,11){\makebox(0,0){\includegraphics{grozin_andrey_fig27.pdf}}}
\put(0,3){\makebox(0,0){$\blue{p_1}$}}
\put(19,3){\makebox(0,0){$\blue{p_2}$}}
\put(32,3){\makebox(0,0){$\green{p'}$}}
\end{picture}}
+ (1\leftrightarrow2)
= \frac{g}{\blue{p_{2-}} \green{p'_+} - i0} + (1\leftrightarrow2)\,.
\end{equation*}
In coordinate space,
\begin{align*}
\red{C_3}(\blue{t_1},\blue{t_2},\green{t'}) &{}= \int \frac{d\blue{p_{1-}}}{2\pi} \frac{d\blue{p_{2-}}}{2\pi} \frac{d\green{p'_+}}{2\pi}
e^{- i \blue{p_{1-} t_1} - i \blue{p_{2-} t2} + i \green{p'_+ t'}}
\red{\tilde{C}_3}(\blue{p_{1-}},\blue{p_{2-}},\green{p'_+})\\
&{}= g \delta(\blue{t_1}) \theta(-\blue{t_2}) \theta(\green{t'}) + (1\leftrightarrow2)\,.
\end{align*}

In other words,
\begin{equation*}
J_3(x) = g\,\blue{\varphi_{c_+}(x)}\,
\int_{-\infty}^0 d\blue{t}\,\blue{\varphi_{c_+}(x + t e_-)}\,
\int_0^\infty d\green{t'}\,\green{\varphi_{c_-}(x + t' e_+)}\,.
\end{equation*}
Incoming momenta $\blue{p_-}$ correspond to the operator $\blue{i\partial_-}$;
outgoing $\green{p'_+}$ -- to $\green{-i\partial_+}$:
\begin{equation*}
J_3(x) = g\,\blue{\varphi_{c_+}(x)}\;
\blue{\frac{1}{i \partial_- - i0} \varphi_{c_+}(x)}\;
\green{\frac{1}{- i \partial_+ - i0} \varphi_{c_-}(x)}\,.
\end{equation*}
Let's check:
\begin{align*}
&\int_{-\infty}^0 dt\,\varphi(x + t e_-)
= \int \frac{d^d k}{(2\pi)^d} e^{-i k\cdot x} \tilde{\varphi}(k) \int_{-\infty}^0 dt\,e^{-i k_- t + 0t}\\
&{} = \int \frac{d^d k}{(2\pi)^d} \frac{i}{k_- + i0} e^{-i k\cdot x} \tilde{\varphi}(k)
= \frac{i}{i\partial_- + i0} \int \frac{d^d k}{(2\pi)^d} e^{-i k\cdot x} \tilde{\varphi}(k)
= \frac{i}{i\partial_- + i0} \varphi(x)\,;
\end{align*}
similarly,
\begin{equation*}
\int_0^\infty dt\,\varphi(x + t e_+) = \frac{-i}{i\partial_+ - i0} \varphi(x)\,.
\end{equation*}

\begin{wrapfigure}[4]{r}{0.3\textwidth}
\vspace{-6mm}
\begin{center}
\begin{picture}(14,14)
\put(7,7){\makebox(0,0){\includegraphics{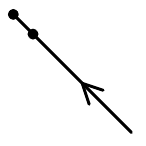}}}
\put(4,12){\makebox(0,0)[l]{$x$}}
\put(2,14){\makebox(0,0)[l]{$x + \alpha e_-$}}
\end{picture}
\end{center}
\caption{$F(x+\alpha e_-)$.}
\label{F:line}
\end{wrapfigure}
\noindent
We can understand this relation also in the following way.
Let a function $F(x)$ be defined as
\begin{equation*}
F(x) = \int_{-\infty}^0 dt\,\varphi(x + t e_-)\,.
\end{equation*}
Then, for an infinitesimal $\alpha$ (Fig.~\ref{F:line}),
\begin{equation*}
F(x + \alpha e_-) = F(x) + \alpha \partial_- F(x)
= \int_{-\infty}^\alpha dt\,\varphi(x + t e_-) = F(x) + \alpha \varphi(x)\,,
\end{equation*}
and hence $\partial_- F(x) = \varphi(x)$\,.

\begin{figure}[h]
\begin{center}
\begin{picture}(104,25)
\put(18,14){\makebox(0,0){\includegraphics{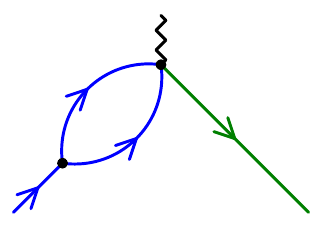}}}
\put(3,6){\makebox(0,0){$\blue{p}$}}
\put(33,8){\makebox(0,0){$\green{p'}$}}
\put(4,16){\makebox(0,0){$\blue{k+p}$}}
\put(18,9){\makebox(0,0){$\blue{-k}$}}
\put(16,0){\makebox(0,0)[b]{a}}
\put(74,14){\makebox(0,0){\includegraphics{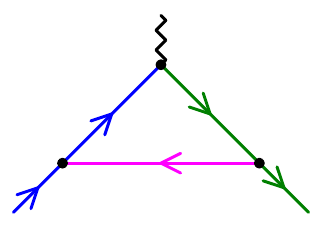}}}
\put(59,6){\makebox(0,0){$\blue{p}$}}
\put(89,8){\makebox(0,0){$\green{p'}$}}
\put(74,6){\makebox(0,0){$\magenta{k}$}}
\put(54,15){\makebox(0,0){$(\magenta{k}_++\blue{p}_+,\blue{p}_-,\blue{\vec{0}})$}}
\put(94,15){\makebox(0,0){$(\green{p'}_+,\magenta{k}_-+\green{p'}_-,\green{\vec{0}})$}}
\put(74,0){\makebox(0,0)[b]{b}}
\end{picture}
\end{center}
\caption{$\blue{c_+}$ collinear (a) and soft (b) loop diagrams in SCET.}
\label{F:FFSCET}
\end{figure}
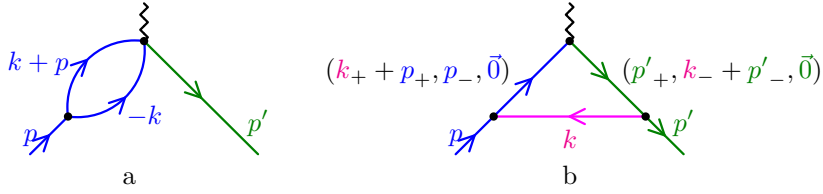

Now we are ready to calculate the form factor
with off-shell momenta $\blue{p = (p_+,p_-,\vec{0})}$, $\green{p' = (p'_+,p'_-,\vec{0})}$
in SCET.
At 1 loop
\begin{equation*}
\raisebox{-5mm}{\begin{picture}(24,14)
\put(12,7){\makebox(0,0){\includegraphics{grozin_andrey_fig24.pdf}}}
\put(2,3){\makebox(0,0){$p$}}
\put(22,3){\makebox(0,0){$p'$}}
\end{picture}}
= \red{\tilde{C}_2^{(1)}} \raisebox{-3mm}{\includegraphics{grozin_andrey_fig23.pdf}}
+ \red{\tilde{C}_3^{(0)}} \raisebox{-5mm}{\includegraphics{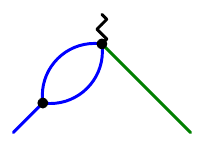}}
+ \red{\tilde{C}_3^{(0)}} \raisebox{-5mm}{\includegraphics{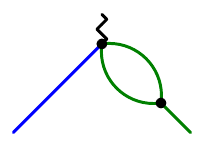}}
+ \raisebox{-5mm}{\includegraphics{grozin_andrey_fig25.pdf}}\,,
\end{equation*}
where $\red{\tilde{C}_2^{(1)}}$ is given by the formula~(\ref{SCET:C2}).
The $\blue{c_+}$ collinear loop diagram (Fig.~\ref{F:FFSCET}a) is
\begin{align*}
&\frac{1}{2} \frac{g^2}{(4\pi)^{d/2}} \int \frac{d^d\blue{k}}{i\pi^{d/2}}
\frac{\red{\tilde{C}_3}(\blue{k_-+p_-},\blue{-k_-},\green{p'_+})}{(\blue{-k^2-i0}) (\blue{-(k+p)^2-i0})}\\
&{} = \frac{g^2}{(4\pi)^{d/2}} \int \frac{d^d\blue{k}}{i\pi^{d/2}}
\frac{1}{(\blue{-k^2-i0}) (\blue{-(k+p)^2-i0}) (- \green{p'}_+ \blue{k}_- -i0)}
= \frac{g^2}{(4\pi)^{d/2}} \blue{I_{c_+}}\,,
\end{align*}
where $\blue{I_{c_+}}$ is given by the formula~(\ref{ScalFF:Ic});
the $\green{c_-}$ collinear loop is similar.
In the soft loop diagram (Fig.~\ref{F:FFSCET}b),
$\blue{p = (p_+,p_-,\vec{0})}$, $\green{p' = (p'_+,p'_-,\vec{0})}$, $\magenta{k = (k_+,k_-,\vec{k}_\bot)}$;
momentum conservation holds up to $\mathcal{O}(\lambda)$, and
\begin{equation*}
\frac{g^2}{(4\pi)^{d/2}} \int \frac{d^d\magenta{k}}{i\pi^{d/2}}
\frac{1}{(\magenta{-k^2-i0}) (\blue{- p_- (k_+ + p_+) - i0}) (\green{- p'_+ (k_- + p'_-) - i0})}
= \frac{g^2}{(4\pi)^{d/2}} \magenta{I_s}\,,
\end{equation*}
where $\magenta{I_s}$ is given by the formula~(\ref{ScalFF:Is}).
Thus we have reproduced the result of the method of regions (Sect.~\ref{S:Reg})
within SCET.

\section{QCD}
\label{S:QCD}

\subsection{SQET Lagrangian}
\label{S:L}

Now we consider massless QCD
\begin{equation*}
L = \bar{\psi} i \D \psi - \frac{1}{4} F^{a\mu\nu} F^a_{\mu\nu}\,,
\end{equation*}
where $i g F^{\mu\nu} = [i D^\mu,i D^\nu]$
and flavor indices are assumed.
We want to construct an effective theory
with soft ($\blue{k_c \sim (\lambda^2,1,\lambda) Q}$)
and collinear ($\magenta{k_s \sim (\lambda^2,\lambda^2,\lambda^2) Q}$)
modes:
\begin{equation*}
\psi \to \blue{\psi_c} + \magenta{\psi_s}\,,\quad
A^\mu \to \blue{A_c^\mu} + \magenta{A_s^\mu}\,.
\end{equation*}
Of course, in any real physical problems there must be some other characteristic momentum in addition to $\blue{k_c}$,
otherwise we cannot say that a nearly light-like momentum $\blue{k_c}$ along $e_+$ has high energy:
either the second nearly light-like momentum along $e_-$ ($\green{k_{c_-}} \sim (q,\lambda^2,\lambda) Q$),
or the heavy-quark momentum $Mv$ ($v=e_0$), or something else.
We'll not consider $\green{c_-}$ collinear modes explicitly;
results for them can be easily obtained by interchanging $+ \leftrightarrow -$.

The correlator of soft quark fields is
\begin{equation*}
\magenta{{<}T \psi_s(x_s) \bar{\psi}_s(0){>}
= \int \frac{d^4 k}{(2\pi)^4} e^{-i k\cdot x} \frac{i \rlap/k}{k^2 + i0} \sim \lambda^6}\,,
\end{equation*}
and hence
\begin{equation*}
\magenta{\psi_s \sim \lambda^3}\,;
\end{equation*}
the correlator of soft gluon fields is
\begin{equation*}
\magenta{{<}T A_s^\mu(x_s) A_s^\nu(0){>}
= \int \frac{d^4 k}{(2\pi)^4} e^{-i k\cdot x} \frac{i}{k^2 + i0}
\left[ - g^{\mu\nu} + \xi \frac{k^\mu k^\nu}{k^2} \right]
\sim \lambda^4}\,,
\end{equation*}
and hence
\begin{equation*}
\magenta{A_s^\mu \sim \lambda^2}
\end{equation*}
(note that the covariant derivative $i D_s^\mu = i \partial_s^\mu + g A_s^\mu \sim \lambda^2$
is homogeneous).
The soft Lagrangian is
\begin{equation*}
\magenta{L_s = \bar{\psi}_s i \D_s \psi_s - \frac{1}{4} F_s^{a\mu\nu} F^a_{s\mu\nu}}\,,
\end{equation*}
where $\magenta{i g F_s^{\mu\nu} = [i D_s^\mu,i D_s^\nu]}$;
it is $\magenta{L_s \sim \lambda^4}$.

The collinear quark field can be decomposed as
\begin{equation}
\blue{\psi_c = \xi + \eta}\,,\quad
\blue{\xi = \frac{\gamma_+ \gamma_-}{4} \psi_c}\,,\quad
\blue{\eta = \frac{\gamma_- \gamma_+}{4} \psi_c}\,.
\label{L:psi}
\end{equation}
The correlators of these fields are
\begin{align*}
&\blue{{<}T \xi(x) \xi(0){>}
= \int \frac{d^4 k}{(2\pi)^4} e^{-i k\cdot x} \frac{i}{k^2 + i0}
\frac{\gamma_+ \gamma_-}{4} \underbrace{\rlap/k}_{\frac{1}{2} k_- \gamma_+} \frac{\gamma_- \gamma_+}{4}
\sim \lambda^2}\,,\\
&\blue{{<}T \eta(x) \eta(0){>}
= \int \frac{d^4 k}{(2\pi)^4} e^{-i k\cdot x} \frac{i}{k^2 + i0}
\frac{\gamma_- \gamma_+}{4} \underbrace{\rlap/k}_{\frac{1}{2} k_+ \gamma_-} \frac{\gamma_+ \gamma_-}{4}
\sim \lambda^4}\,,
\end{align*}
and hence
\begin{equation*}
\blue{\xi \sim \lambda}\,,\quad
\blue{\eta \sim \lambda^2}\,.
\end{equation*}
The correlator of collinear gluon fields is
\begin{equation*}
\blue{{<}T A_c^\mu(x) A_c^\nu(0){>}
= \int \frac{d^4 k}{(2\pi)^4} e^{-i k\cdot x} \frac{i}{k^2 + i0}
\left[ - g^{\mu\nu} + \xi \frac{k^\mu k^\nu}{k^2} \right]}
\end{equation*}
($\xi$ is the gauge parameter), and hence
\begin{equation*}
\blue{{<}A_{c+} A_{c+}{>} \sim \lambda^4}\,,\quad
\blue{{<}A_{c-} A_{c-}{>} \sim 1}\,,\quad
\blue{{<}A_{c\bot} A_{c\bot}{>} \sim \lambda^2}\,,
\end{equation*}
and $\blue{A_c \sim (\lambda^2,1,\lambda)}$
(note that the covariant derivative
$\blue{i D_c^\mu = i \partial_c^\mu + g A_c^\mu \sim (\lambda^2,1,\lambda)}$
is homogeneous).

Substituting the decomposition~(\ref{L:psi}) into the QCD Lagrangian we obtain
\begin{align*}
\blue{L} &\blue{{}= i (\bar{\xi} + \bar{\eta})
\left( \frac{1}{2} D_+ \gamma_- + \frac{1}{2} D_- \gamma_+ + \D_\bot \right)
(\xi + \eta)}\,,\\
&\blue{{}= \frac{i}{2} \bar{\xi} D_+ \gamma_- \xi
+ \frac{i}{2} \bar{\eta} D_- \gamma_+ \eta
+ i \bar{\xi} \D_\bot \eta
+ i \bar{\eta} \D_\bot \xi}\,.
\end{align*}
The equation of motion
\begin{equation*}
\blue{\D \psi_c =
\left( \frac{1}{2} D_+ \gamma_- + \frac{1}{2} D_- \gamma_+ + \D_\bot \right)
(\xi + \eta) = 0}
\end{equation*}
is decomposed into
\begin{equation*}
\blue{\gamma_+ \D \psi_c = 2 D_+ \xi + \D_\bot \eta = 0}\,,\quad
\blue{\gamma_- \D \psi_c = 2 D_- \eta + \D_\bot \xi = 0}\,.
\end{equation*}
We can express the small field $\eta$ via the leading field $\xi$:
\begin{equation*}
\blue{\eta = - \frac{1}{2} \gamma_- \frac{1}{i D_- + i0} i \D_\bot \xi}\,,\quad
\blue{\bar{\eta} = - \frac{1}{2} \bar{\xi} i \rlap{\hspace{1mm}/}\overleftarrow{D}_\bot
\frac{1}{i \overleftarrow{D}_- + i0} \gamma_-}\,.
\end{equation*}
Substituting this solution into the Lagrangian
(or, equivalently, integrating out $\eta$)
leads to
\begin{align*}
\blue{L ={}}& \blue{\frac{1}{2} \bar{\xi} i D_+ \gamma_- \xi
- \frac{1}{2} \bar{\xi} i \D_\bot \frac{\gamma_-}{i D_- + i0} i \D_\bot \xi
- \frac{1}{2} \bar{\xi} i \rlap{\hspace{1mm}/}\overleftarrow{D}_\bot
\frac{\gamma_-}{i \overleftarrow{D}_- + i0} i \D_\bot \xi}\\
&\blue{{} + \frac{1}{8} \bar{\xi} i \rlap{\hspace{1mm}/}\overleftarrow{D}_\bot
\frac{\gamma_-}{i \overleftarrow{D}_- + i0} i D_- \gamma_+ \frac{\gamma_-}{D_-} \D_\bot \xi}\\
\blue{{}={}}& \blue{\frac{1}{2} \bar{\xi} i D_+ \gamma_- \xi
+ \frac{1}{2} \bar{\xi} i \D_\bot \frac{1}{i D_- + i0} i \D_\bot \gamma_- \xi}\,.
\end{align*}

This Lagrangian contains different orders in $\lambda$.
In order to obtain the leading-order Lagrangian we substitute
$i D_- \to \blue{i D_{c-} = i \partial_- + g A_{c-}}$,
$i D_\bot \to \blue{i D_{c\bot} = i \partial_\bot + g A_{c\bot}}$
and arrive at
\begin{align}
&L = \magenta{\bar{\psi}_s i \D_s \psi_s
- \frac{1}{4} F^{a\mu\nu}_s F^a_{s\mu\nu}}
\nonumber\\
&{} + \blue{\frac{1}{2} \bar{\xi} \biggl[
i \underbrace{D_+}_{i \partial_+ + g A_{c+} + g \magenta{A_{s+}(\bar{x}_+)}}
+ i \D_{c\bot} \frac{1}{i D_{c-} + i0} i \D_{c\bot} \biggr] \gamma_- \xi}
- \blue{\frac{1}{4} F^{a\mu\nu}_c F^a_{c\mu\nu}}\,,
\label{L:SCET}
\end{align}
where $\magenta{\bar{x}_+ \equiv \frac{1}{2} x_- e_+^\mu}$ (multipole expansion)
and $i g \blue{F_c^{\mu\nu}} = [i D^\mu,i D^\nu]$,
$i D^\mu = \frac{1}{2} i D_+ e_-^\mu + \frac{1}{2} i \blue{D_{c-}} e_+^\mu + i \blue{D_{c\bot}^\mu}$.

\begin{sloppypar}
The SCET Lagrangian~(\ref{L:SCET}) is invariant with respect to soft gauge transformations
$\magenta{U_s(x) = e^{i f_s^a(x) t^a}}$
where the characteristic scale of $\magenta{U_s}$
is $\magenta{x_s \sim (\lambda^{-2},\lambda^{-2},\lambda^{-2})}$:
\begin{equation*}
\magenta{\psi_s(x) \to U_s(x) \psi_s(x)}\,,\quad
\magenta{A_s^\mu(x) \to U_s(x) A_s^\mu(x) U_s^+(x) - \frac{i}{g} \bigl(\partial^\mu U_s(x)\bigr) U_s^{-1}(x)}\,.
\end{equation*}
The collinear fields transform as
\begin{equation*}
\blue{\xi(x)} \to \magenta{U_s(\bar{x}_+)} \blue{\xi(x)}\,,\quad
\blue{A_c^\mu(x)} \to \magenta{U_s(\bar{x}_+)} \blue{A_c^\mu(x)} \magenta{U_s^{-1}(\bar{x}_-)}\,,
\end{equation*}
because
\begin{equation*}
\magenta{U_s}(\blue{x}) =
\magenta{U_s}(\blue{\bar{x}_+})
- \underbrace{\blue{\vec{x}_\bot}\cdot\magenta{\vec{\partial}_\bot U_s}(\blue{\bar{x}_+})}_{\mathcal{O}(\lambda)}
+ \mathcal{O}(\lambda^2)\,.
\end{equation*}
The term $\magenta{\bigl(\partial^\mu U_s\bigr) U_s^{-1}} \sim \lambda^2$
is not needed for $\blue{A_{c-}}$ and $\blue{A_{c\bot}}$.
The component $\blue{A_{c+}}$ only appears in $D_+$:
\begin{equation*}
\blue{A_{c+}(x)} + \magenta{A_{s+}(\bar{x}_+)} \to
\magenta{U_s(\bar{x}_+)}
\left[ \blue{A_{c+}(x)} + \magenta{A_{s+}(\bar{x}_+)} \right]
\magenta{U_s^{-1}(\bar{x}_+)}
+ \frac{i}{g} \magenta{U_s(\bar{x}_+)}
\bigl(\partial_+ \magenta{U_s^{-1}(\bar{x}_+)} \bigr)\,,
\end{equation*}
and $i D_+ \to \magenta{U_s(\bar{x}_+)} i D_+ \magenta{U_s^{-1}(\bar{x}_+)}$.
\end{sloppypar}

The Lagrangian~(\ref{L:SCET}) is also invariant with respect to collinear gauge transformations
$\blue{U_c(x) = e^{i f_c^a(x) t^a}}$
where the characteristic scale of $\blue{U_c}$
is $\blue{x_c \sim (1,\lambda^{-2},\lambda^{-1})}$.
These fast varying transformation cannot affect smooth soft fields:
\begin{equation*}
\magenta{\psi_s(x)} \to \magenta{\psi_s(x)}\,,\quad
\magenta{A_s^\mu(x)} \to \magenta{A_s^\mu(x)}\,.
\end{equation*}
The collinear fields transform as
\begin{align*}
&\blue{\xi(x)} \to \blue{U_c(x) \xi(x)}\,,\\
&\blue{A_c^\mu(x)} \to \blue{U_c(x) A_c^\mu(x) U_c^{-1}(x)}
+ \frac{1}{g} \blue{U_c(x)} \left[i \partial^\mu + \tfrac{1}{2} g \magenta{A_{s+}(\bar{x}_+)} e_-^\mu,\blue{U_c^{-1}(x)}\right]\,.
\end{align*}
For components we have
\begin{align*}
&\blue{A_{c\bot}(x)} \to \blue{U_c(x) A_{c\bot}(x) U_c^{-1}(x)}
+ \frac{i}{g} \blue{U_c(x) \left(\partial_\bot U_c^{-1}(x)\right)}\,,\\
&\blue{A_{c-}(x)} \to \blue{U_c(x) A_{c-}(x) U_c^{-1}(x)}
+ \frac{i}{g} \blue{U_c(x) \left(\partial_- U_c^{-1}(x)\right)}\,,\\
&\blue{A_{c+}(x)} \to \blue{U_c(x) A_{c+}(x) U_c^{-1}(x)}
+ \frac{i}{g} \blue{U_c(x)} \left[\magenta{D_{s+}(\bar{x}_+)},\blue{U_c^{-1}(x)}\right]\,,
\end{align*}
so that $i D_+ \to \blue{U_c(x)} i D_+ \blue{U_c^{-1}(x)}$.

The SCET Lagrangian~(\ref{L:SCET}) is not Lorentz invariant --
it explicitly contains the vectors $e_+$, $e_-$.
However, these vectors are not uniquely defined by the physical problem.
We may change them keeping soft momenta soft,
and collinear momenta -- collinear.
All physical predictions remain intact
(reparametrization invariance).
This symmetry is generated by the following infinitesimal transformations:
\begin{itemize}
\item Lorentz boost $e_+ \to (1+\varepsilon) e_+$, $e_- \to (1-\varepsilon) e_-$:
collinear momenta remain collinear
($\blue{k_+} \to (1+\varepsilon) \blue{k_+}$,
$\blue{k_-} \to (1-\varepsilon) \blue{k_-}$);
\item $e_+ \to e_+$, $e_- \to e_- + \varepsilon_\bot$
(where $\varepsilon_\bot$ is an infinitesimal transverse vector):
\begin{equation*}
\blue{k_-} \to \underbrace{\blue{k_-}}_{\mathcal{O}(1)}
+ \underbrace{\blue{k}\cdot\varepsilon_\bot}_{\mathcal{O}(\lambda)}\,,\quad
\blue{k_\bot} = \blue{k} - \tfrac{1}{2} (\blue{k}\cdot e_+) e_- - \tfrac{1}{2} (\blue{k}\cdot e_-) e_+
\to \underbrace{\blue{k_\bot}}_{\mathcal{O}(\lambda)}
- \underbrace{\tfrac{1}{2} \blue{k_+} \varepsilon_\bot}_{\mathcal{O}(\lambda^2)}
- \underbrace{\tfrac{1}{2} (\blue{k}\cdot\varepsilon_\bot) e_+}_{\mathcal{O}(\lambda)}\,;
\end{equation*}
\item $e_+ \to e_+ + \lambda \varepsilon_\bot$, $e_- \to e_-$:
\begin{equation*}
\blue{k_+} \to \underbrace{\blue{k_+}}_{\mathcal{O}(\lambda^2)}
+ \underbrace{\lambda \blue{k}\cdot\varepsilon_\bot}_{\mathcal{O}(\lambda^2)}\,,\quad
\blue{k_\bot} \to \underbrace{\blue{k_\bot}}_{\mathcal{O}(\lambda)}
- \underbrace{\tfrac{\lambda}{2} (\blue{k}\cdot\varepsilon_\bot) e_-}_{\mathcal{O}(\lambda^2)}
- \underbrace{\tfrac{\lambda}{2} \blue{k_-} \varepsilon_\bot}_{\mathcal{O}(\lambda)}
\end{equation*}
(this transformation mixes different orders in $\lambda$).
\end{itemize}
Of course, if there are also $\green{c_-}$ collinear modes,
the second transformation also has to contain $\lambda \varepsilon_\bot$,
to keep $\green{c_-}$ collinear momenta $\green{c_-}$ collinear.

\begin{wrapfigure}[9]{r}{0.3\textwidth}
\vspace{-5mm}
\begin{center}
\begin{picture}(14,26)
\put(7,13){\makebox(0,0){\includegraphics{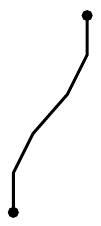}}}
\put(6.25,3){\makebox(0,0){$x$}}
\put(7.75,23){\makebox(0,0){$x'$}}
\end{picture}
\end{center}
\caption{Wilson line.}
\label{F:Wilson}
\end{wrapfigure}
Wilson lines are widely used in SCET.
Therefore we make a short digression about their properties.
The Wilson line in QCD is defined as
\begin{equation}
[x',x] = P\exp i g \int_x^{x'} dy_\mu\,A^\mu(y)\,;
\label{L:Wilson}
\end{equation}
it depends on a contour connecting $x$ with $x'$.
Any contour can be represented by a sequence of infinitesimal straight segments (Fig.~\ref{F:Wilson});
the path-ordered exponent in~(\ref{L:Wilson}) is defined as the product of the infinitesimal Wilson lines,
corresponding to these segments (from $x$ to $x'$), arranged from right to left.
An infinitesimal Wilson line transforms under a gauge transformation as
\begin{align*}
&[x+dx,x] = 1 + i g A^\mu(x)\,dx_\mu \to
1 + i g \left[ U(x) A^\mu(x) U^{-1}(x) - \frac{i}{g} \left(\partial^\mu U(x)\right) U^{-1}(x) \right] dx_\mu\\
&{} = U(x+dx) U^{-1}(x) + i g U(x) A^\mu(x) U^{-1}(x)\,dx_\mu
= U(x+dx) \left[1 + i g A^\mu(x)\,dx_\mu\right] U^{-1}(x)\,,
\end{align*}

\begin{wrapfigure}{r}{0.3\textwidth}
\vspace{-5mm}
\begin{center}
\begin{picture}(18,28)
\put(9,13){\makebox(0,0){\includegraphics{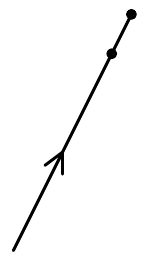}}}
\put(10.5,21){\makebox(0,0){$x$}}
\put(6,25){\makebox(0,0){$x+v\,dt$}}
\end{picture}
\end{center}
\caption{$W(x+v\,dt)$.}
\label{F:Wdx}
\end{wrapfigure}
\noindent
and hence for any finite Wilson line we obtain
\begin{equation*}
[x',x] \to U(x') [x',x] U^{-1}(x)\,.
\end{equation*}
This means that, e.\,g., $\bar{\psi}(x') [x',x] \psi(x)$ is a gauge invariant operator.
Let's define
\begin{equation*}
W(x) = [x,x-\infty v] = P \exp i g \int_{-\infty}^0 dt\,v_\mu A^\mu(x+vt)\,.
\end{equation*}
If we consider only gauge transformations which are identical in the infinite past
($U(x-\infty v) = 1$), then $W(x) \to U(x) W(x)$,
and $W^{-1}(x) \psi(x)$ (as well as $\bar{\psi}(x) W(x)$) is a gauge invariant operator.
Let's extend this line (Fig.~\ref{F:Wdx}): $W(x+v\,dt) = \left[1 + i g v\cdot A(x)\,dt \right] W(x)$,
and hence $v\cdot\partial W(x) = i g v\cdot A(x)\,W(x)$, or $v\cdot D\,W(x) = 0$.

Collinear quarks interact with soft gluons:
\begin{equation*}
L_q = \frac{1}{2} \blue{\bar{\xi}(x)} i D_+ \gamma_- \blue{\xi(x)}\,,\quad
i D_+ = i \partial_+ + g \blue{A_{c+}(x)} + g \magenta{A_{s+}(\bar{x}_+)}
= i \blue{D_{c+}} + g \magenta{A_{s+}(\bar{x}_+)}\,.
\end{equation*}

\begin{wrapfigure}{r}{0.3\textwidth}
\vspace{-15mm}
\begin{center}
\begin{picture}(42,42)
\put(21,21){\makebox(0,0){\includegraphics{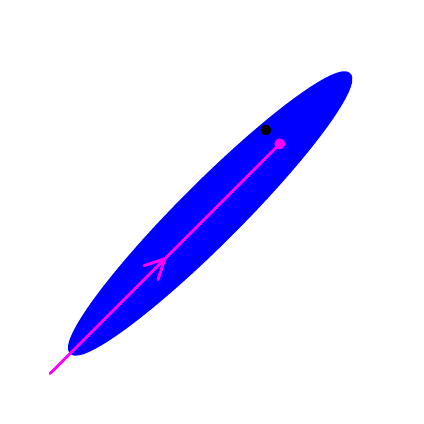}}}
\put(25.57,30.57){\makebox(0,0){$x$}}
\put(30.57,25.57){\makebox(0,0){$\magenta{\bar{x}_+}$}}
\end{picture}
\end{center}
\vspace{-10mm}
\caption{Soft Wilson line.}
\label{F:Ws}
\end{wrapfigure}
\noindent
If we introduce the soft Wilson line (Fig.~\ref{F:Ws})
\begin{equation*}
\magenta{S(x) = P \exp i g \int_{-\infty}^0 dt\, A_{s+}(x + e_+ t)}
\end{equation*}
(where $D_{s+} S(x) = 0$)
and redefine the collinear fields as
\begin{equation*}
\blue{\xi(x)} = \magenta{S(\bar{x}_+)} \blue{\xi_0(x)}\,,\quad
\blue{A_c^\mu(x)} = \magenta{S(\bar{x}_+)} \blue{A_{c0}^\mu(x)} \magenta{S^{-1}(\bar{x}_+)}\,,
\end{equation*}
then
\begin{align*}
&i D_+ \blue{\xi(x)}\\
&{} = \bigl[ \magenta{S(\bar{x}_+)} i \partial_+
+ \bigl(\magenta{\bigl(i \partial_+ + g A_s(\bar{x}_+) \bigr) S(\bar{x}_+)}\bigr)
+ g \magenta{S(\bar{x}_+)} \blue{A_{c0+}(x)} \bigr] \blue{\xi_0(x)}
= \magenta{S(\bar{x}_+)} \blue{i D_{c0+} \xi_0(x)}\,,
\end{align*}
and the Lagrangian becomes
\begin{equation*}
L_q = \frac{1}{2} \blue{\bar{\xi_0}(x) i D_{c0+} \xi_0(x)}\,.
\end{equation*}
So, in terms of the new field $\blue{\xi_0(x)}$ there is no interaction with soft gluons.

Similarly, collinear gluons interact with soft gluons:
$\blue{F_c^{\mu\nu}}$ in the Lagrangian~(\ref{L:SCET}) contains not only $\blue{A_c^\mu}$
but also $\magenta{A_s^\mu}$ (because of $D_+$).
The transformation to $\blue{A_{c0}^\mu}$ leads to
\begin{equation*}
i \blue{D_{c-}} = i \partial_- + g \magenta{S(\bar{x}_+)} \blue{A_{c0-}(x)} \magenta{S^{-1}(\bar{x}_+)}
= \magenta{S(\bar{x}_+)} i \blue{D_{c0-}} \magenta{S^{-1}(\bar{x}_+)}\,,
\end{equation*}
because $\partial_- \magenta{S(\bar{x}_+)} = 2 \partial\magenta{S(\bar{x}_+)}/\partial x_+ = 0$;
$i \blue{D_{c\bot}} = \magenta{S(\bar{x}_+)} i \blue{D_{c0\bot}} \magenta{S^{-1}(\bar{x}_+)}$;
and
\begin{align*}
&i D_+ = \magenta{S(\bar{x}_+)} i \partial_+ \magenta{S^{-1}(\bar{x}_+)}
- \magenta{S(\bar{x}_+)} \bigl(i \partial_+ \magenta{S^{-1}(\bar{x}_+)}\bigr)
+ g \magenta{A_{s+}(\bar{x}_+)}
+ g \magenta{S(\bar{x}_+)} \blue{A_{c0+}(x)} \magenta{S^{-1}(\bar{x}_+)}\\
&{} = \magenta{S(\bar{x}_+)} \blue{i D_{c0+}} \magenta{S^{-1}(\bar{x}_+)}
\end{align*}
because $\magenta{S(\bar{x}_+)} \bigl(i \partial_+ \magenta{S^{-1}(\bar{x}_+)}\bigr)
= - \bigl(i \partial_+ \magenta{S(\bar{x}_+)}\bigr) \magenta{S^{-1}(\bar{x}_+)}
= g \magenta{A_{s+}(\bar{x}_+)}$,
$\magenta{i D_{s+} S = (i \partial_+ + g A_{s+}) S = 0}$
(note that $\blue{i D_{c0+} = i \partial_+ + g A_{c0+}}$ does not contain the soft gluon field).
Therefore,
\begin{equation*}
L_g = - \frac{1}{2} \Tr \blue{F_{c0\mu\nu} F_{c0}^{\mu\nu}},
\end{equation*}
where $\blue{F_c^{\mu\nu}(x)} = \magenta{S(\bar{x}_+)} \blue{F_{c0}^{\mu\nu}(x)} \magenta{S^{-1}(\bar{x}_+)}$,
and $\blue{F_{c0}^{\mu\nu}}$ does not contain the soft gluon field.
So, in terms of the new field $\blue{A_{c0}^\mu}$ there is no interaction with soft gluons.

\begin{wrapfigure}{r}{0.3\textwidth}
\vspace{-15mm}
\begin{center}
\begin{picture}(42,42)
\put(21,21){\makebox(0,0){\includegraphics{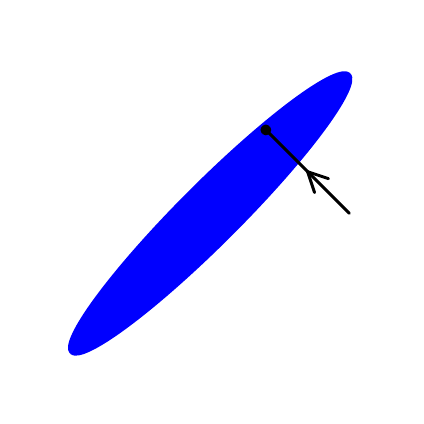}}}
\put(25.57,30.57){\makebox(0,0){$x$}}
\end{picture}
\end{center}
\vspace{-10mm}
\caption{Collinear Wilson line.}
\label{F:Wc}
\end{wrapfigure}
Using the collinear Wilson line (Fig.~\ref{F:Wc})
\begin{equation*}
\blue{W(x)} = P \exp i g \int_{-\infty}^0 dt\,\blue{A_{c-}(x + t e_-)}
\end{equation*}
we can redefine the collinear fields as
\begin{equation*}
\blue{\xi(x)} = \blue{W(x)} \blue{\chi(x)}\,,\quad
\blue{\mathcal{A}^\mu(x)} = \blue{W^{-1}(x)} \bigl(i \blue{D_c^\mu W(x)}\bigr)
\end{equation*}
(note that $\blue{\mathcal{A}}_- = 0$, $\blue{\mathcal{A}}_+ \sim \lambda^2$, $\blue{\mathcal{A}}_\bot \sim \lambda$).
Defining also the covariant derivative
\begin{equation*}
i \blue{\mathcal{D}^\mu} = \blue{W^{-1}(x) i D_c^\mu W(x)}
= i \partial^\mu + \blue{\mathcal{A}^\mu},
\end{equation*}
we can rewrite the collinear quark Lagrangian in~(\ref{L:SCET}) as
\begin{equation*}
\blue{L_q} = \frac{1}{2} \blue{\bar{\chi} i \mathcal{D}_+ \gamma_- \chi}
+ \frac{1}{2} \blue{\bar{\chi} i \DD_\bot} \frac{1}{i \partial_- + i0} \blue{i \DD_\bot \gamma_- \chi}
= \frac{1}{2} \blue{\bar{\chi} i \mathcal{D}_+ \gamma_- \chi}
- \frac{i}{2} \blue{\bar{\chi} \gamma_- i \DD_\bot}
\int_{-\infty}^0 dt\,\left(\blue{i \DD_\bot \chi}\right)_{x + t e_-}\,,
\end{equation*}
because $\blue{W^{-1}(x) i D_{c-} W(x)} = i \partial_-$ and hence
\begin{equation*}
\blue{W^{-1}(x) \frac{1}{i D_{c-} + i0} W(x)} = \frac{1}{i \partial_- + i0}\,.
\end{equation*}
Nonlocality in $e_-$ direction is $\sim Q^{-1}$.
Similarly, the collinear gluon Lagrangian in~(\ref{L:SCET}) becomes
\begin{equation*}
\blue{L_g} = - \frac{1}{2 g^2} \Tr \blue{\mathcal{F}_{\mu\nu} \mathcal{F}^{\mu\nu}}\,,
\end{equation*}
where
\begin{equation*}
\blue{W^{-1} F_{c\mu\nu} W = \mathcal{F}_{\mu\nu}}
= \frac{1}{g} \blue{\left( \partial_\mu \mathcal{A}_\nu - \partial_\nu \mathcal{A}_\mu
- i [\mathcal{A}_\mu,\mathcal{A}_\nu]\right)}\,.
\end{equation*}

\subsection{Vector quark current}
\label{S:J}

Now we shall consider the quark vector current
\begin{equation*}
J^\mu(x) = \bar{\psi}(x) \gamma^\mu \psi(x)
\end{equation*}
in the kinematical situation $\green{c_-} \to \blue{c_+}$.
Of course, we need both $\blue{c_+}$ collinear fields (discussed in Sect.~\ref{S:L})
and $\green{c_-}$ collinear ones
(all formulas for them can be easily obtained by the interchange $+ \leftrightarrow -$).
The QCD field $\psi(x)$ becomes $\magenta{S_-(\bar{x}_-)} \green{\chi_-(x)}$
(where $\bar{x}_- = \frac{1}{2} x_+ e_-$, $\gamma_- \green{\chi_-} = 0$);
similarly, $\bar{\psi}(x)$ becomes $\blue{\bar{\chi}_+(x)} \magenta{S^+_+(\bar{x}_+)}$
(where $\bar{x}_+ = \frac{1}{2} x_- e_+$, $\gamma_+ \blue{\chi_+} = 0$).
Similarly to Sect.~\ref{S:SCET},
\begin{equation*}
J^\mu(x) = \int d\blue{t}\,d\green{t'}\,\red{C_V}(\blue{t},\green{t'}) \blue{\bar{\chi}_+(x + t e_-)}
\magenta{S_+^+(\bar{x}_+) S_-(\bar{x}_-)}
\gamma_\bot^\mu \green{\chi_-(x + t' e_+)} + \cdots
\end{equation*}
The integration region is $x^\mu \sim (1,1,\lambda^{-1}) Q^{-1}$ (Fig.~\ref{F:xcoll}),
and we can use the multipole expansion to get
\begin{equation*}
J^\mu(x) = \int d\blue{t}\,d\green{t'}\,\red{C_V}(\blue{t},\green{t'}) \blue{\bar{\chi}_+(\bar{x}_- + x_\bot + t e_-)}
\magenta{S_+^+(0) S_-(0)}
\gamma_\bot^\mu \green{\chi_-(\bar{x}_+ + x_\bot + t' e_+)} + \cdots
\end{equation*}

So, at the leading order in $\lambda$ the matrix element of $J^\mu$ (the form factor) can be factorized,
\begin{equation*}
\raisebox{-10mm}{\begin{picture}(26,22)
\put(13,11){\makebox(0,0){\includegraphics{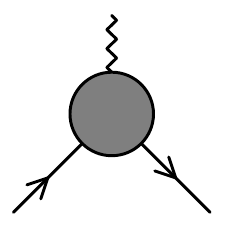}}}
\put(1.5,1.5){\makebox(0,0){$p$}}
\put(24.5,1.5){\makebox(0,0){$p'$}}
\end{picture}} =
\raisebox{-14mm}{\begin{picture}(40,31)
\put(20,15.5){\makebox(0,0){\includegraphics{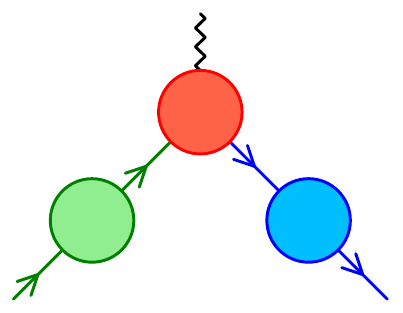}}}
\end{picture}} \otimes
\raisebox{-8mm}{\begin{picture}(32,17)
\put(16,8.5){\makebox(0,0){\includegraphics{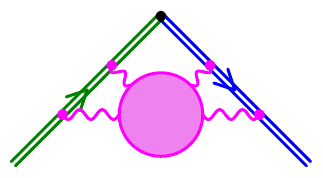}}}
\end{picture}}\,,
\end{equation*}
to all orders in $\alpha_s$.
The matrix element of $\magenta{S_+^+(0) S_-(0)}$ is the soft factor $\magenta{S}$,
the soft Wilson line with the cusp.
The matching coefficient $\red{\tilde{C}_V}$ is the hard factor,
and the propagators of $\blue{\chi_+}$ and $\green{\chi_-}$ produce the jet factors:
\begin{equation*}
F(-q^2,-p^2,-p^{\prime2}) = \red{\tilde{C}_V(-q^2)} \green{J_-(-p^2)} \blue{J_+(-p^{\prime2})} \magenta{S\left(\frac{(-p^2) (-p^{\prime2})}{-q^2}\right)}\,.
\end{equation*}

The coefficient $\red{C_V}$ is obtained by matching the on-shell matrix elements in QCD and SCET:
the quark momenta are $\green{p = (p_+,0,\vec{0})}$, $\blue{p'=(0,p'_-,\vec{0})}$.
At the tree level
\begin{equation*}
\raisebox{-5mm}{\begin{picture}(14,11)
\put(7,5.5){\makebox(0,0){\includegraphics{grozin_andrey_fig22.pdf}}}
\put(1,4){\makebox(0,0){$\green{p}$}}
\put(13,4){\makebox(0,0){$\blue{p'}$}}
\end{picture}}
= \red{\tilde{C}_V}(\green{p_+},\blue{p'_-})
\raisebox{-5mm}{\begin{picture}(14,11)\put(7,5.5){\makebox(0,0){\includegraphics{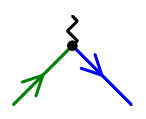}}}\end{picture}}\,,
\end{equation*}
and hence
\begin{equation*}
\red{\tilde{C}_V^{(0)}}(\green{p_+},\blue{p'_-}) = 1\,,\quad
\red{C_V^{(0)}}(\green{t},\blue{t'}) = \delta(\green{t}) \delta(\blue{t'})\,.
\end{equation*}
At 1 loop
\begin{equation*}
\raisebox{-5mm}{\begin{picture}(24,14)
\put(12,7){\makebox(0,0){\includegraphics{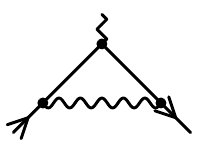}}}
\put(2,3){\makebox(0,0){$\green{p}$}}
\put(22,3){\makebox(0,0){$\blue{p'}$}}
\end{picture}}
= \red{\tilde{C}_V^{(1)}}(\green{p_+},\blue{p'_-})
\raisebox{-5mm}{\begin{picture}(14,11)\put(7,5.5){\makebox(0,0){\includegraphics{grozin_andrey_fig42.pdf}}}\end{picture}}\,,
\end{equation*}
because SCET loops vanish.
The calculation of this loop diagram is similar to that of $\red{I_h}$ (Sect.~\ref{S:ScalReg}),
the result is
\begin{align*}
\red{C_V(-q^2)} &{}= 1 - C_F \frac{g_0^2 (-q^2)^{-\varepsilon}}{(4\pi)^{d/2}}
\frac{\Gamma(1+\varepsilon) \Gamma^2(1-\varepsilon)}{\Gamma(1-2\varepsilon)}
\left[ \frac{2}{\varepsilon^2} + \frac{3+2\varepsilon}{\varepsilon (1-2\varepsilon)} \right]\\
&{} = 1 - C_F \frac{g_0^2 (-q^2)^{-\varepsilon}}{(4\pi)^{d/2}} e^{-\gamma\varepsilon}
\left( \frac{2}{\varepsilon^2} + \frac{3}{\varepsilon} - \frac{\pi^2}{6} + 8 + \mathcal{O}(\varepsilon) \right)\,.
\end{align*}
Re-expressing this matching coefficient via the $\overline{\text{MS}}$ coupling $\alpha_s(\mu)$,
we obtain
\begin{equation*}
\red{C_V(-q^2)} = 1 - C_F \frac{\alpha_s(\mu)}{4\pi} \left(\frac{-q^2}{\mu^2}\right)^{-\varepsilon}
\left( \frac{2}{\varepsilon^2} + \frac{3}{\varepsilon} - \frac{\pi^2}{6} + 8 \right)\,.
\end{equation*}
The bare matching coefficient should be equal to $\red{C_V(-q^2)} = \red{Z^{-1} C_V(-q^2,\mu)}$,
where the renormalized matching coefficient $\red{C_V(-q^2,\mu)}$ is finite at $\varepsilon\to0$,
and $\red{Z}$ is a minimal renormalization constant (containing only negative powers of $\varepsilon$).
We arrive at
\begin{align*}
&\red{Z} = 1 + C_F \left( \frac{2}{\varepsilon^2} - \frac{2}{\varepsilon} \log\frac{-q^2}{\mu^2} + \frac{3}{\varepsilon} \right)
\frac{\alpha_s(\mu)}{4\pi}\,,\\
&\red{C_V(-q^2,\mu)} = 1 + C_F \left( - \log^2\frac{-q^2}{\mu^2} + 3 \log\frac{-q^2}{\mu^2} + \frac{\pi^2}{6} - 8 \right)
\frac{\alpha_s(\mu)}{4\pi}\,.
\end{align*}

The bare matching coefficient does not depend on $\mu$:
\begin{equation*}
\frac{d\log\red{C_V(-q^2)}}{d\log\mu} = 0 = - \frac{d\log\red{Z}}{d\log\mu} + \frac{d\log\red{C_V(-q^2,\mu)}}{d\log\mu}\,.
\end{equation*}
The SCET vector current anomalous dimension is defined by
\begin{equation*}
\frac{d\log\red{Z}}{d\log\mu} = \Gamma(\alpha_s(\mu)) \log\frac{-q^2}{\mu^2} + \red{\gamma_V(\alpha_s(\mu))}\,;
\end{equation*}
the renormalized matching coefficient satisfies the renormalization group (RG) equation
\begin{equation}
\frac{d\log\red{C_V(-q^2,\mu)}}{d\log\mu} = \Gamma(\alpha_s(\mu)) \log\frac{-q^2}{\mu^2} + \red{\gamma_V(\alpha_s(\mu))}\,.
\label{J:RG}
\end{equation}
Taking into account $d\log\alpha_s(\mu)/d\log\mu = - 2 \varepsilon + \mathcal{O}(\alpha_s)$,
we obtain
\begin{equation*}
\Gamma(\alpha_s) = 4 C_F \frac{\alpha_s}{4\pi} + \mathcal{O}(\alpha_s^2)\,,\quad
\red{\gamma_V(\alpha_s)} = - 6 C_F \frac{\alpha_s}{4\pi} + \mathcal{O}(\alpha_s^2)\,.
\end{equation*}
Here $\Gamma(\alpha_s)$ is called the light-like cusp anomalous dimension.

The solution of the RG equation~(\ref{J:RG}) can be written as
\begin{equation*}
\red{C_V(-q^2,\mu)} = U(\mu_0,\mu) \red{C_V(-q^2,\mu_0)}\,,
\end{equation*}
where the evolution factor $U(\mu_0,\mu)$ satisfies the equation
\begin{equation*}
\frac{d\log U(\mu_0,\mu)}{d\log\mu} = \Gamma(\alpha_s(\mu)) \log\frac{-q^2}{\mu^2} + \gamma_V(\alpha_s(\mu))
\end{equation*}
with the initial condition $U(\mu_0,\mu_0) = 1$.
The anomalous dimensions have the form
\begin{equation*}
\Gamma(\alpha_s) = \Gamma_0 \frac{\alpha_s}{4\pi} + \Gamma_1 \left(\frac{\alpha_s}{4\pi}\right)^2 + \cdots\,,\quad
\gamma_V(\alpha_s) = \gamma_{V0} \frac{\alpha_s}{4\pi} + \gamma_{V1} \left(\frac{\alpha_s}{4\pi}\right)^2 + \cdots\,.
\end{equation*}
Dividing this equation by the evolution equation for $\alpha_s(\mu)$,
\begin{equation*}
\frac{d\log\alpha_s(\mu)}{d\log\mu} = - 2 \beta(\alpha_s(\mu))\,,\quad
\beta(\alpha_s) = \beta_0 \frac{\alpha_s}{4\pi} + \beta_1 \left(\frac{\alpha_s}{4\pi}\right)^2 + \cdots\,,
\end{equation*}
we obtain
\begin{equation*}
\frac{d\log U(\mu_0,\mu)}{d\log\alpha_s} = - \frac{1}{2 \beta(\alpha_s)}
\left[ \Gamma(\alpha_s) \left( \log\frac{-q^2}{\mu_0^2} - 2 \log\frac{\mu}{\mu_0} \right) + \gamma_V(\alpha_s) \right]\,.
\end{equation*}
We substitute
\begin{equation*}
\log\frac{\mu}{\mu_0} = - \int_{\alpha_s(\mu_0)}^{\alpha_s(\mu)} \frac{d\alpha_s}{\alpha_s} \frac{1}{2 \beta(\alpha_s)}
\end{equation*}
and arrive at the solution
\begin{equation}
U(\mu_0,\mu) = \exp \left[ S(\mu_0,\mu) - A_{\gamma_V}(\mu_0,\mu) \right]
\left(\frac{-q^2}{\mu_0^2}\right)^{-A_\Gamma(\mu_0,\mu)}\,,
\label{J:sol}
\end{equation}
where
\begin{align*}
&A_\gamma(\mu_0,\mu) = \int_{\alpha_s(\mu_0)}^{\alpha_s(\mu)} \frac{d\alpha_s}{\alpha_s} \frac{\gamma(\alpha_s)}{2 \beta(\alpha_s)}
= \frac{\gamma_0}{2 \beta_0} \log r + \mathcal{O}(\alpha_s)\,,\\
&S(\mu_0,\mu) = - \int_{\alpha_s(\mu_0)}^{\alpha_s(\mu)} \frac{d\alpha_s}{\alpha_s} \frac{\Gamma(\alpha_s)}{2 \beta(\alpha_s)}
\int_{\alpha_s(\mu_0)}^{\alpha_s} \frac{d\alpha_s'}{\alpha_s'} \frac{1}{\beta(\alpha_s')}\\
&{} = \frac{\Gamma_0}{2 \beta_0^2} \biggl[
\frac{4\pi}{\alpha_s(\mu_0)} \left( \frac{r-1}{r} - \log r \right)
+ \left( \frac{\Gamma_1}{\Gamma_0} - \frac{\beta_1}{\beta_0} \right) \left( 1 - r + \log r \right)
+ \frac{\beta_1}{2 \beta_0} \log^2 r \biggr] + \mathcal{O}(\alpha_s)\,,
\end{align*}
and $r = \alpha_s(\mu)/\alpha_s(\mu_0)$.

\begin{wrapfigure}{r}{0.2\textwidth}
\vspace{-10mm}
\begin{center}
\begin{picture}(40,22)
\put(20,11){\makebox(0,0){\includegraphics{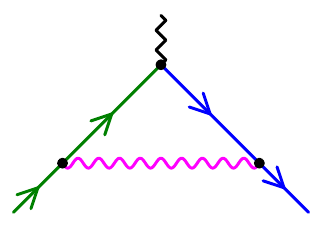}}}
\put(4,3.5){\makebox(0,0){$p$}}
\put(36,3.5){\makebox(0,0){$p'$}}
\put(7,11){\makebox(0,0){$k+p$}}
\put(33,11){\makebox(0,0){$k+p'$}}
\put(20,4){\makebox(0,0){$k$}}
\put(22,18.5){\makebox(0,0){$q$}}
\end{picture}
\end{center}
\vspace{-5mm}
\caption{Soft function.}
\label{F:S}
\end{wrapfigure}
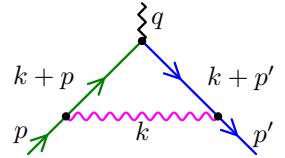
The soft function at 1 loop (Fig.~\ref{F:S}) is
\begin{equation*}
\magenta{S^{(1)}} = - 2 C_F \frac{g_0^2}{(4\pi)^{d/2}} (-q^2) \magenta{I_s}\,,
\end{equation*}
see Sect.~\ref{S:ScalReg}.
Therefore,
\begin{align*}
&\magenta{S(\Lambda_s^2)} = 1 - 2 C_F \frac{g_0^2 (\Lambda_s^2)^{-\varepsilon}}{(4\pi)^{d/2}} \Gamma(1-\varepsilon) \Gamma^2(\varepsilon)\\
&{} = 1 - 2 C_F \frac{\alpha_s(\mu)}{4\pi} \left(\frac{\Lambda_s^2}{\mu^2}\right)^{-\varepsilon}
\left( \frac{1}{\varepsilon^2} + \frac{\pi^2}{4} \right)
= \magenta{Z_S^{-1} S(\Lambda_s^2,\mu)}\,,
\end{align*}
where
\begin{align*}
&\magenta{Z_S} = 1 + 2 C_F \left( \frac{1}{\varepsilon^2} - \frac{1}{\varepsilon} \log\frac{\Lambda_s^2}{\mu^2} \right) \frac{\alpha_s(\mu)}{4\pi}
+ \mathcal{O}(\alpha_s^2)\,,\\
&\magenta{S(\Lambda_s^2,\mu)} = 1 - C_F \left( \log^2\frac{\Lambda_s^2}{\mu^2} + \frac{\pi^2}{2} \right) \frac{\alpha_s(\mu)}{4\pi}
+ \mathcal{O}(\alpha_s^2)\,.
\end{align*}
The renormalized soft function satisfies the RG equation
\begin{equation*}
\frac{d\log\magenta{S(\Lambda_s^2,\mu)}}{d\log\mu} = \Gamma(\alpha_s(\mu)) \log\frac{\Lambda_s^2}{\mu^2} + \magenta{\gamma_S(\alpha_s(\mu))}\,,
\end{equation*}
where
\begin{equation*}
\magenta{\gamma_S(\alpha_s)} = 0 + \mathcal{O}(\alpha_s^2)\,.
\end{equation*}

To summarize: the quark form factor at the leading order in $\lambda$ is equal to
\begin{equation}
F(-q^2,-p^2,-p^{\prime2}) = \red{C_V(-q^2,\mu)} \green{J(-p^2,\mu)} \blue{J(-p^{\prime2},\mu)} \magenta{S(\Lambda_s^2,\mu)}\,,
\label{J:Fact}
\end{equation}
where
\begin{equation*}
\magenta{\Lambda_s^2} = \frac{(-p^2) (-p^{\prime2})}{-q^2}\,.
\end{equation*}
It does not depend on $\mu$:
\begin{align*}
&\frac{d\log F(-q^2,-p^2,-p^{\prime2})}{d\log\mu} = 0\\
&{} = \frac{d\log\red{C_V(-q^2,\mu)}}{d\log\mu}
+ \frac{d\log\green{J(-p^2,\mu)}}{d\log\mu}
+ \frac{d\log\blue{J(-p^{\prime2},\mu)}}{d\log\mu}
+ \frac{d\log\magenta{S(\Lambda_s^2,\mu)}}{d\log\mu}\,.
\end{align*}
The anomalous dimensions are
\begin{align*}
&\frac{d\log\red{C_V(-q^2,\mu)}}{d\log\mu} = \Gamma(\alpha_s(\mu)) \log\frac{-q^2}{\mu^2} + \gamma_V(\alpha_s(\mu))\,,\\
&\frac{d\log\green{J(-p^2,\mu)}}{d\log\mu} = - \Gamma(\alpha_s(\mu)) \log\frac{-p^2}{\mu^2} - \gamma_J(\alpha_s(\mu))\,,\\
&\frac{d\log\magenta{S(\Lambda_s^2,\mu)}}{d\log\mu} = \Gamma(\alpha_s(\mu)) \log\frac{\Lambda_s^2}{\mu^2} + \gamma_S(\alpha_s(\mu))\,,
\end{align*}
where $\Gamma(\alpha_s)$ is the light-like cusp anomalous dimension.
Therefore, the following condition has to be satisfied:
\begin{align*}
&\Gamma(\alpha_s(\mu)) \biggl[ \log\frac{-q^2}{\mu^2} - \log\frac{-p^2}{\mu^2} - \log\frac{-p^{\prime^2}}{\mu^2}
+ \log\frac{(-p^2) (-p^{\prime2})}{(-q^2) \mu^2} \biggr]\\
&{} + \gamma_V(\alpha_s(\mu)) - 2 \gamma_J(\alpha_s(\mu)) + \gamma_S(\alpha_s(\mu)) = 0\,.
\end{align*}
This condition explains why the coefficient of the logarithm in each anomalous dimension is $\Gamma(\alpha_s)$.

\begin{figure}[h]
\begin{center}
\begin{picture}(77,42)
\put(43.5,21){\makebox(0,0){\includegraphics{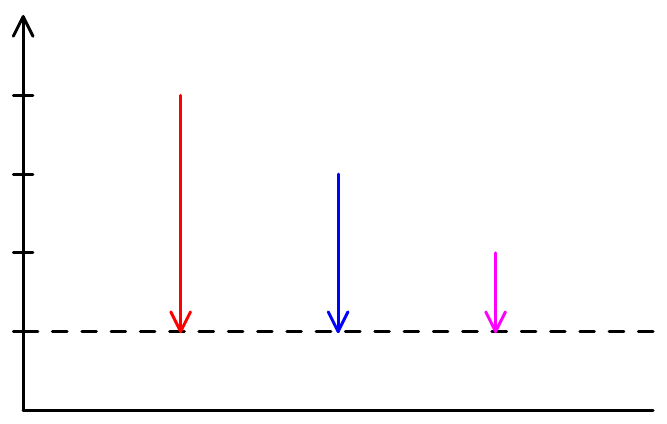}}}
\put(10,33){\makebox(0,0)[r]{$-q^2$}}
\put(10,25){\makebox(0,0)[r]{$-p^2 \sim -p^{\prime2}$}}
\put(10,17){\makebox(0,0)[r]{$\Lambda_s^2$}}
\put(10,9){\makebox(0,0)[r]{$\mu^2$}}
\put(28,34){\makebox(0,0)[b]{$\red{C_V(-q^2,\mu)}$}}
\put(44,26){\makebox(0,0)[b]{$\blue{J(-p^2,\mu)}$}}
\put(60,18){\makebox(0,0)[b]{$\magenta{S(\Lambda^2,\mu)}$}}
\end{picture}
\end{center}
\caption{Running the factors in the factorization formula.}
\label{F:run}
\end{figure}
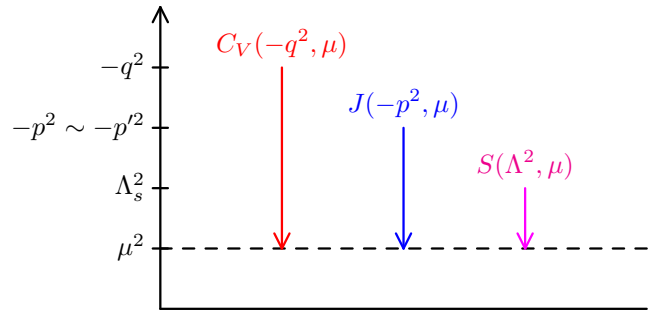

We have to use the same normalization scale $\mu$ in all factors in the factorization formula~(\ref{J:Fact}).
However, in order to avoid large logarithms in perturbative series
it is better to calculate $\red{C_V(-q^2,\mu)}$ at $\mu^2 \sim -q^2$;
$\green{J(-p^2,\mu)}$ and $\blue{J(-p^{\prime2},\mu)}$ at $\mu^2 \sim -p^2 \sim -p^{\prime2}$;
and $\magenta{S(\Lambda_s^2,\mu)}$ at $\mu^2 \sim \Lambda_s^2$.
Then we use the RG equations to run these factors to a common $\mu^2$ (Fig.~\ref{F:run}).

\section{Conclusion}
\label{S:Conc}

In these lectures we discussed basics of SCET-I as applied to a single problem,
the form factor with large $-q^2$ and small $-p^2 \sim -p^{\prime2}$.
For applications of SCET to numerous physical problems,
both in heavy quark physics and in high energy hadron collider physics,
see~\cite{BBF:15,BS} and references cited therein.

I am grateful to the organizers of the Helmholtz--DIAS International Summer School
``Quantum Field Theory at the Limits: from Strong Fields to Heavy Quarks''
(Dubna, July 18--30) for inviting me.
The work was supported by the Russian ministry of education and science.

\end{document}